%% file: main.tex
\documentclass[acmtog]{acmart}
\acmSubmissionID{1123}
\usepackage{cleveref}
\usepackage{amsmath}
\usepackage{xcolor}
\usepackage{colortbl}
\usepackage{url}

\usepackage[ruled]{algorithm2e} 

\SetAlFnt{\small}
\SetAlCapFnt{\small}
\SetAlCapNameFnt{\small}
\SetAlCapHSkip{0pt}

\input{src_macros}

\begin{document}
\setcopyright{cc}
\setcctype{by}
\acmJournal{TOG}
\acmYear{2026} \acmVolume{45} \acmNumber{4}
\acmMonth{7} \acmDOI{10.1145/3811369}
\acmArticle{62}

\title{Gabor Fields: Orientation-Selective Level-of-Detail for Volume Rendering}

\begin{teaserfigure}
  \centering
  \includegraphics[width=\linewidth]{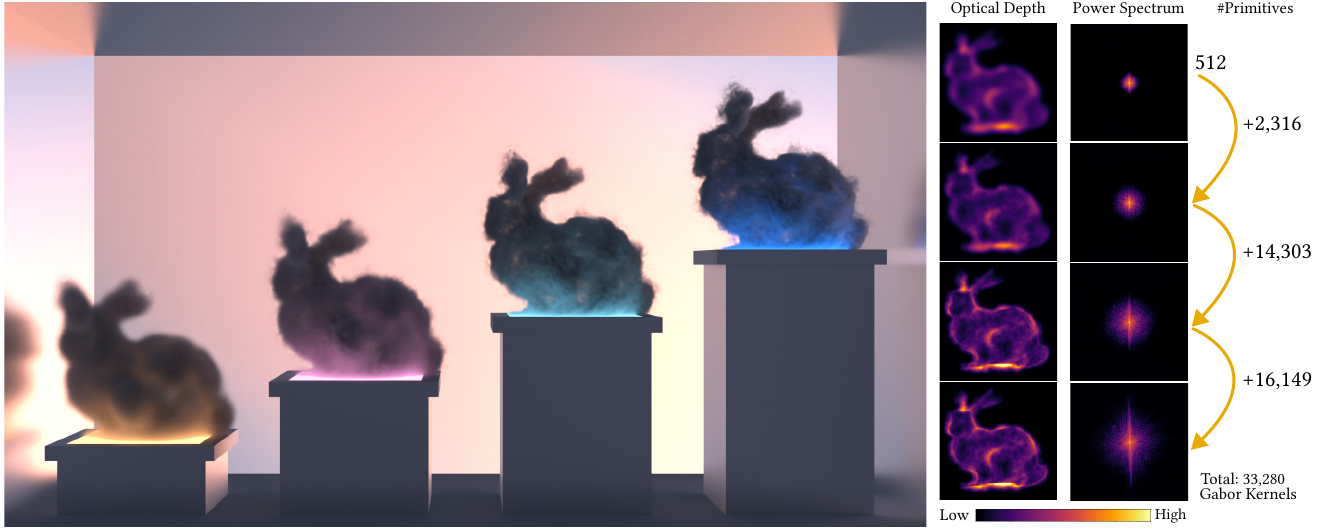}
  \caption{Example level-of-detail decomposition of our Gabor Fields, a new volumetric density field representation using primitive Gabor kernels. The frequency- and orientation-selectivity of the representation enables continuous control over level of detail, reduced rendering time, and a trade-off between variance and bias for performance when sample cost is critical. Here, we show four levels obtained by masking particle collections from the highest-quality asset at render time. (Volumetric {\sc{}Bunny} from the OpenVDB database~\cite{openvdb})
  }
  \label{fig:teaser}
  \Description[Teaser Bunny]{Bunny}
\end{teaserfigure}
\author{Jorge Condor}
\authornote{Both authors contributed equally to this research.}
\email{jorge.condor@usi.ch}
\orcid{0000-0002-9958-0118}
\affiliation{%
  \institution{Università della Svizzera Italiana, Lugano}
  \streetaddress{Via Giuseppe Buffi 13}
  \city{Lugano}
  \state{Ticino}
  \country{Switzerland}
  \postcode{6900}
}%
\author{Nicolai Hermann}
\authornotemark[1]
\email{nicolai.hermann@usi.ch}
\orcid{0009-0004-7936-6757}
\affiliation{%
  \institution{Università della Svizzera Italiana, Lugano}
  \streetaddress{Via Giuseppe Buffi 13}
  \city{Lugano}
  \country{Switzerland}
  \postcode{6900}
}%
\author{Mehmet Ata Yurtsever}
\email{ata.yurtsever@usi.ch}
\orcid{0009-0006-6509-760X}
\affiliation{%
  \institution{Università della Svizzera Italiana, Lugano}
  \streetaddress{Via Giuseppe Buffi 13}
  \city{Lugano}
  \country{Switzerland}
  \postcode{6900}
}%
\author{Piotr Didyk}
\email{piotr.didyk@usi.ch}
\orcid{0000-0003-0768-8939}
\affiliation{%
  \institution{Università della Svizzera Italiana, Lugano}
  \streetaddress{Via Giuseppe Buffi 13}
  \city{Lugano}
  \country{Switzerland}
  \postcode{6900}
}%
\renewcommand{\shortauthors}{Condor, et al.}
\begin{abstract}
\input{sections_paper/00_abstract}
\end{abstract}


\maketitle

\input{sections_paper/01_introduction}
\input{sections_paper/02_related_work}
\input{sections_paper/03_background}
\input{sections_paper/04_our_method}
\input{sections_paper/05_regression}
\input{sections_paper/06_results}

\input{sections_paper/07_applications}
\input{sections_paper/08_limitations_future_work}
\input{sections_paper/09_conclusions}

\bibliographystyle{ACM-Reference-Format}
\bibliography{bibliography}
\appendix
\input{sections_paper/A_appendix}

\input{sections_paper/B_supplementary}

\end{document}

%% file: src_macros.tex
\definecolor{lightred}{rgb}{1,0.4,0.4}
\definecolor{lightyellow}{rgb}{1,1,0.4}
\definecolor{lightorange}{rgb}{1,0.647,0.4}

\newcommand{\point}[1]{\mathbf{#1}}
\newcommand{\px}{\point{x}}

\newcommand{\sRad}{L}

\newcommand{\sTrans}{\textrm{T}}

%% file: sections_paper/00_abstract.tex
Gaussian-based representations have enabled efficient physically-based volume rendering at a fraction of the memory cost of regular, discrete, voxel-based distributions. One of the remaining advantages of classic voxel grids, however, is the ease of constructing hierarchical representations by either storing volumetric mipmaps or selectively pruning branches of an already hierarchical voxel grid. Such strategies reduce rendering time and eliminate aliasing when lower levels of detail are required. Constructing similar strategies for Gaussian-based volumes is not trivial. Straightforward solutions, such as prefiltering or computing mipmap-style representations, lead to increased memory requirements or expensive re-fitting of each level separately. Additionally, such solutions do not guarantee a smooth transition between different hierarchy levels. To address these limitations, we propose Gabor Fields, a mixture of Gabor kernels that enables continuous, orientation-selective frequency filtering at no cost. The frequency content of the asset is reduced by selectively pruning primitives, directly benefiting rendering performance, without refitting or extra storage. Beyond filtering, we demonstrate that stochastically sampling from different frequencies and orientations at each ray recursion enables masking substantial portions of the volume, accelerating ray traversal time in single- and multiple-scattering settings. Furthermore, inspired by procedural volumes, we present an application for efficient design and rendering of procedural clouds as Gabor-noise-modulated Gaussians. Code available here \url{https://arcanous98.github.io/projectPages/gaborVolumes.html}.

%% file: sections_paper/01_introduction.tex
\section{Introduction}\label{sec:intro}
High-quality volumetric representations are central for many applications in computer graphics. They are widely used, especially when obtaining or using explicit representations is difficult or prohibitively expensive. Examples include modeling and simulation of particle aggregates~\cite{meng15, moon07}, clouds~\cite{kallweit2017}, foliage~\cite{loubet17, neyret98}, cloth~\cite{shroder11, zhao11} and hair~\cite{petrovic-2005}. Most of the above methods use discrete density distributions as their primitive model, organized in 3D voxel grids that can be dense, adaptive~\cite{museth2013vdb} or partly implicit~\cite{neuralvdb}. A significant drawback of these methods is their poor memory scalability and the need to use stochastic methods for efficient rendering with physically-based path tracers~\cite{novak2018monte}, which limits the potential of volumetric rendering techniques. 

In recent years, Lagrangian-style particle methods have become the volumetric representation of choice underpinning applications including radiance field reconstructions \cite{lassner-2021,kerbl20233Dgaussians, wu-2024, zhou2025splat}, simultaneous localization and mapping (SLAM) \cite{keetha-2024}, mesh extraction \cite{guedon-2024}, or fast tomographic rendering~\cite{gao-2025}. Inspired by these developments, Condor et al.~\shortcite{Condor2024Gaussians} proposed a technique to efficiently model and render kernel mixture models bounded by volumetric primitives. The method replaced discrete volumetric density distributions with continuous ones, providing substantial memory and speed improvements over traditional voxel grids. The generality of this technique enables both path tracing and radiance field rendering, offering multiple advantages over previous rasterized frameworks, such as support for complex camera models, the ability to trace secondary rays to compute relighting, shadowing, scattering, and analytic transmittance integrals. However, the approach has several limitations. The modeling power of the Gaussian kernels is limited, and the regression scales poorly. Moreover, Gaussian volumetric primitives are not directly compatible with level-of-detail (LOD) strategies. The analogous approach to downsampling voxel grids would require convolving primitives with a low-pass filter, for example, another Gaussian, producing new Gaussian primitives with larger spatial supports. Such a representation would have the same complexity yet slower rendering time. While aggregating primitives after filtering is possible, it would require energy-based optimizations or re-fitting of the asset. 

To address these limitations, we draw inspiration from steerable pyramid decompositions~\cite{Simoncelli1992,Simoncelli1995} and propose a hierarchical volumetric primitive representation based on mixtures of Gaussian and Gabor kernels. Gabor kernels, harmonically modulated Gaussians, have a frequency-domain representation consisting of symmetric, opposed Gaussians centered around their peak modulation frequency (see Figure~\ref{fig:selectivity_spectral}). This spectral localization makes them ideal for constructing a spectrally aware volumetric representation in which filtering specific spatial frequency bands corresponds to removing certain kernels. This enables a simple LOD strategy that does not require additional memory or create discontinuities between levels. These properties make LOD both natural and efficient, similar to a continuous Laplacian pyramid but with orientation selectivity. We support this representation with analytically derived line integrals for the Gabor kernels and a novel hierarchical regression scheme. In addition, we propose an efficient stochastic rendering approach that leverages the unique properties of the representation to drastically reduce cost per sample in brute-force multiple-scattering path tracing settings. Furthermore, Gabor kernels are orientation-selective. They model residuals only in specific frequencies and orientations, contributing negligibly in others. We demonstrate, for the first time, that this property can be effectively leveraged to improve rendering time by selectively masking particles that do not contribute significantly to a given ray.
Our main contributions can be summarized as:
\begin{itemize}
    \item a novel frequency- and orientation-selective volumetric density representation,
    \item analytic line integral and sampling solutions for tomographic rendering, single and multiple scattering, forward and inverse path tracing,
    \item a regression scheme for distilling 3D density fields into Gabor Fields,
    \item strategies leveraging the new representation to accelerate physically based rendering,
    \item applications to efficient LOD, motion blur, and noise-inspired procedural asset generation.
\end{itemize}

%% file: sections_paper/02_related_work.tex
\section{Related Work}
\label{sec:related_work}
\paragraph{Monte Carlo Volume Rendering}
Simulating light transport in participating media has been a longstanding area of research in computer graphics, typically requiring solutions to the radiative transfer equation~\cite{chandrasekhar1960radiative}. Due to the dimensionality of the integral, it is most often estimated stochastically using Monte Carlo approaches; an excellent survey can be found in~\cite{novak2018monte}. Some popular approaches include ray marching~\cite{tuy1984direct, kettunen21}, also used transversally in many implicit radiance field rendering applications~\cite{mildenhall2020, mueller2022instant}; and null-scattering estimators~\cite{woodcock}, either fully analog~\cite{novak2018monte} or with partial closed-form estimation and sample reweighting (i.e. control variates)~\cite{novak14, szirmay2017unbiased, crespo2021primary}, or leveraging power-series expansions~\cite{georgiev19}. Representing heterogeneous media typically involves voxel grids due to their simplicity and support for hierarchical acceleration structures~\cite{museth2013vdb, museth2021nanovdb}, though these become memory-intensive when high amounts of detail are required. 

\paragraph{Physically Based Volume Rendering with Volumetric Primitives}
Alternatives to voxel grids have been studied in the past. Sparse representations, such as collections of isotropic volumes used in particle physics~\cite{max1979atomlll, Brown03}, suffer from the limited modelling power of isotropic primitives. Recent similar efforts in vision and graphics do not fully leverage the advantages of closed-form integration~\cite{lassner-2021, Knoll2021} and do not support scattering. Other alternatives include hierarchical grids~\cite{ProductionVolumeRendering2017}, implicit MLP mixtures~\cite{Lombardi21, kilonerf} and sparse feature grids~\cite{mueller2022instant}; these trade off quality, speed or compression, but overall do not provide a completely rounded solution. More recently, and inspired by primitive-based radiance field efforts~\cite{kerbl20233Dgaussians}, anisotropic ellipsoid-bounded kernel mixture models~\cite{Condor2024Gaussians, zhou2024unifiedgaussianprimitivesscene} were proposed, providing a compact, fast and relatively high-quality alternative. Our work is directly based on these, where our novel hierarchical, spectrally-bounded regression and the introduction of a new kernel (anisotropic Gabors) improves on some of the limitations of the approach, namely, regression quality for low mixture complexities, rendering speed and efficient support of level of detail.

\paragraph{Gabor Kernels and Noise in Rendering}
A Gabor kernel combines a Gaussian envelope with a sinusoidal wave. Its sensitivity to frequency and orientation is widely exploited in image processing and computer vision. Gabor kernels are commonly used in the study and modeling of human visual perception, as they can mimic the human visual system's sensitivity to specific orientations, spatial frequencies, and locations within the visual field \cite{Marcelja1980}.
They have been successfully used to define procedural noise and texture synthesis methods \cite{lagae2009procedural,filtering2011Lagae,galerne12}, offering a compact and resolution-independent approach to generate structured randomness in textures and materials which reduces storage and bandwidth requirements of image-based textures, and support infinite detail through frequency-aware evaluation. In the particular case of volume modeling and rendering, procedural noise is ubiquitous, as it is common for games and VFX production to allow artists or fluid dynamics solvers to create a guide distribution (i.e., metaballs or constant density ellipsoids~\cite{Wrenninge2011}) and layer noise on top to create compact and visually appealing assets~\cite{fajardo2023stochastic}. Our work bridges the gap between procedural and explicitly generated assets through a unified framework. It enables us to construct extremely compact multi-resolution volumetric fields. Assets can be produced in a forward (procedural noise generation) or inverse manner (regression of already existing discrete density fields), and both can be efficiently rendered leveraging closed-form transmittance integrals and distance sampling, which we pioneer for Gabor kernels in this work. 

\paragraph{Level of Detail in Rendering}

The concept of LOD is crucial for optimizing runtime, memory usage and preventing aliasing. LOD is most commonly associated with meshes, where the geometric complexity of rendered objects is adjusted according to their size in the image space, either through precomputed mesh decompositions~\cite{Luebke2002, surace2023b} or through virtualized geometry~\cite{karis2021nanite}. For appearance, mip-mapping \cite{Crow1984} is a standard LOD technique for textures, where multiple prefiltered versions of a texture, so-called (Gaussian) pyramids, are stored, allowing for quick access to the appropriate level based on the texture's screen-space footprint during rendering. Higher-order appearance aggregation has also been explored, e.g. aggregating local light transport~\cite{kallweit2017,vicini2021non,bako23,weier23,zhou2025sceneagn}. The concept of LOD has also been applied to explicit volumes. One of the pioneering methods for pyramidal volumetric representations, introduced by Laur and Hanrahan \cite{Laur1991}, allowed efficient rendering based on the viewer's location by approximating projected density through kernel splatting; however, it was not possible to compute physically-based light transport on such a structure. Building on these ideas, Weiler et al. \shortcite{Weiler2000} proposed an approach for volumes that ensures consistent interpolation between different resolution levels. Additionally, while not always explicitly, more recent volumetric representations, such as OpenVDB~\cite{museth2013vdb} and derivative works \cite{Museth2021, neuralvdb}, also provide a hierarchy to accelerate volume rendering. In contrast to these methods, our approach does not rely on a discrete volumetric representation; and furthermore, enables continuous control over frequency content, without the need of static, discrete pyramid decompositions. We represent the volume using mixtures of explicit functions, specifically Gabor kernels. 

Closer to our work are wavelet-based representations~\cite{Guthe2002}. Compared to them, our method offers a more expressive representation due to the flexibility of the kernels (which are of arbitrary frequency, covariance and location, as opposed to fixed basis) and analytical transmittance integrals, which are critical for efficient volume rendering.

\paragraph{Level of Detail for Primitive-based Models}

In the context of primitive representations, LOD has been left largely unexplored until very recently. In radiance field reconstructions, work has focused on Gaussian-pyramid-like decompositions: training scenes at different scales, then proposing solutions to online switching between the different levels to maximize performance while minimizing artifacts during rendering~\cite{windisch2025lodgaussiansunifiedtraining}. If explicit storage is unfeasible due to the scale of the asset, chunking systems have been proposed~\cite{kerbl2024hierarchical3dgaussianrepresentation}, as well as streaming from global memory and implicit encoding of Gaussian parameters~\cite{Yang_2025}. Similarly, different texturing alternatives have been proposed in order to compress the representation (via providing higher modelling power per primitive)~\cite{chao2025texturedgaussiansenhanced3d, condor2026nht}, but so far it’s unproven whether the learned textures can be safely filtered to provide LOD tools.

In contrast, we propose the first pure primitive-based volumetric model that can naturally decompose not only into different frequency bands, but that is also orientation-selective. This structure maximizes rendering speed and compression, as we do not need to further compute or store new pyramid levels to provide LOD, but rather simply mask out parts of the same asset to naturally obtain a continuous representation, spectrally-wise.

%% file: sections_paper/03_background.tex
\section{Background - Primitive Volume Rendering}
\label{sec:background}
\input{tables/math_var_table}

Here we summarize how to perform physically-based volume rendering with \emph{volumetric primitives}~\cite{Condor2024Gaussians}. We further compiled a table of the most important variables for reference in Table~\ref{tab:var_name}. Each primitive defines a localized region of matter characterized by a density kernel $K_i(\mathbf{x})$, weight $\alpha_i$, emission, and phase function (the latter two omitted here for conciseness). Each kernel has limited support, being bounded by an ellipsoid with a primary axis radius of $3\sigma$, which in practice and for efficiency reasons, is approximated by a triangle mesh. The extinction field at any point $\mathbf{x}$ is expressed as the sum over contributing primitives:
\begin{equation}
    \kappa_t = \sum_{i=1}^{N}\alpha_i K_i(\mathbf{x})
\end{equation}%
where $N$ is the number of primitives influencing $\mathbf{x}$. {For an arbitrary ray $r = x_0 + \vec{v}\cdot{}t$, each primitive’s} contribution to the optical depth along a ray segment $[a, b]$ is given by
\begin{equation}
     \tau_i (\mathbf{x}_a, \mathbf{x}_b) = \alpha_i \int_a^b K_i(\mathbf{x}_t) dt
\end{equation}
Using an exponential decay model, the total transmittance $T$ through the segment becomes
\begin{equation}
    T(\mathbf{x}_a, \mathbf{x}_b) = \exp\left(-\sum_{i = 1}^N \tau_i(\mathbf{x}_a, \mathbf{x}_b)\right),
\end{equation}

\paragraph{Segment-Based Radiative Transfer.}
To evaluate radiance, the ray is partitioned into segments, determined by entry and exit points of the primitives. Radiance is computed as the sum of each segment's contribution, weighted by the accumulated transmittance:
\begin{equation}
    \sRad(\px_0,\px_{M},\vec{v}) = \sum_{k=1}^M \sTrans_{k-1}(\px_0,\px_k) \sRad_k(\px_{k-1},\px_{k},\vec{v}), 
\end{equation} where \(\sRad_k\) denotes the integrated in-scattered radiance within segment \(k\), and \(\sTrans_{k-1}\) is the cumulative transmittance up to segment \(k\). Sampling the interaction distance along the ray is achieved via inversion of the cumulative density function along the ray, for a given random sample \(\xi \in (0,1)\):

\begin{equation}
    log(1-\xi) = - \sum_{i}^N \tau_i(\mathbf{x_0}, \mathbf{x}_t).
\end{equation}%
This allows for efficient sampling strategies via root-finding (e.g. Newton-Raphson, bisection) or closed-form inversion, depending on the choice of kernels. For dense scenes, uniform sampling within segments offers a bias-performance trade-off that remains effective in practice. 

%% file: tables/math_var_table.tex
\begin{table}[t]
  \centering
  \small
  \caption{Important variables used in the paper with their symbols. }
  
  \begin{tabular}{>{\raggedright\arraybackslash}p{0.15\linewidth}>{\raggedleft\arraybackslash}p{0.75\linewidth}}
    \toprule
    \textbf{Symbol} & \textbf{Variable Name}\\
    \midrule
    $\kappa$ & Extinction Field \\
    $\alpha$ & Kernel Weight \\
    $K_i$ & Kernel \\
    $\tau$ & Optical Depth \\
    $T$ & Transmittance \\
    $L$ & In-scattered Radiance \\
    $\sigma$ & Standard Deviation\\
    \midrule
    $\mu$ & Kernel Mean \\
    $\vec{\omega}$ &  Frequency and Orientation of the Gabor Modulation \\
    $f_0$ & Peak Frequency of the Gabor Modulation \\
    $\Sigma$ & Covariance Matrix \\
    \midrule
    $W$ & Whitening Matrix \\
    $\vec{k}_W$ & Frequency and Orientation in Whitened Space \\
    \bottomrule
  \end{tabular}
  \label{tab:var_name}
\end{table}

%% file: sections_paper/04_our_method.tex
\section{Our method}
\label{sec:our_method}
\subsection{Motivation}
\label{sec:spectral_analysis}
\begin{figure*}
    \centering
    \includegraphics[width=\linewidth]{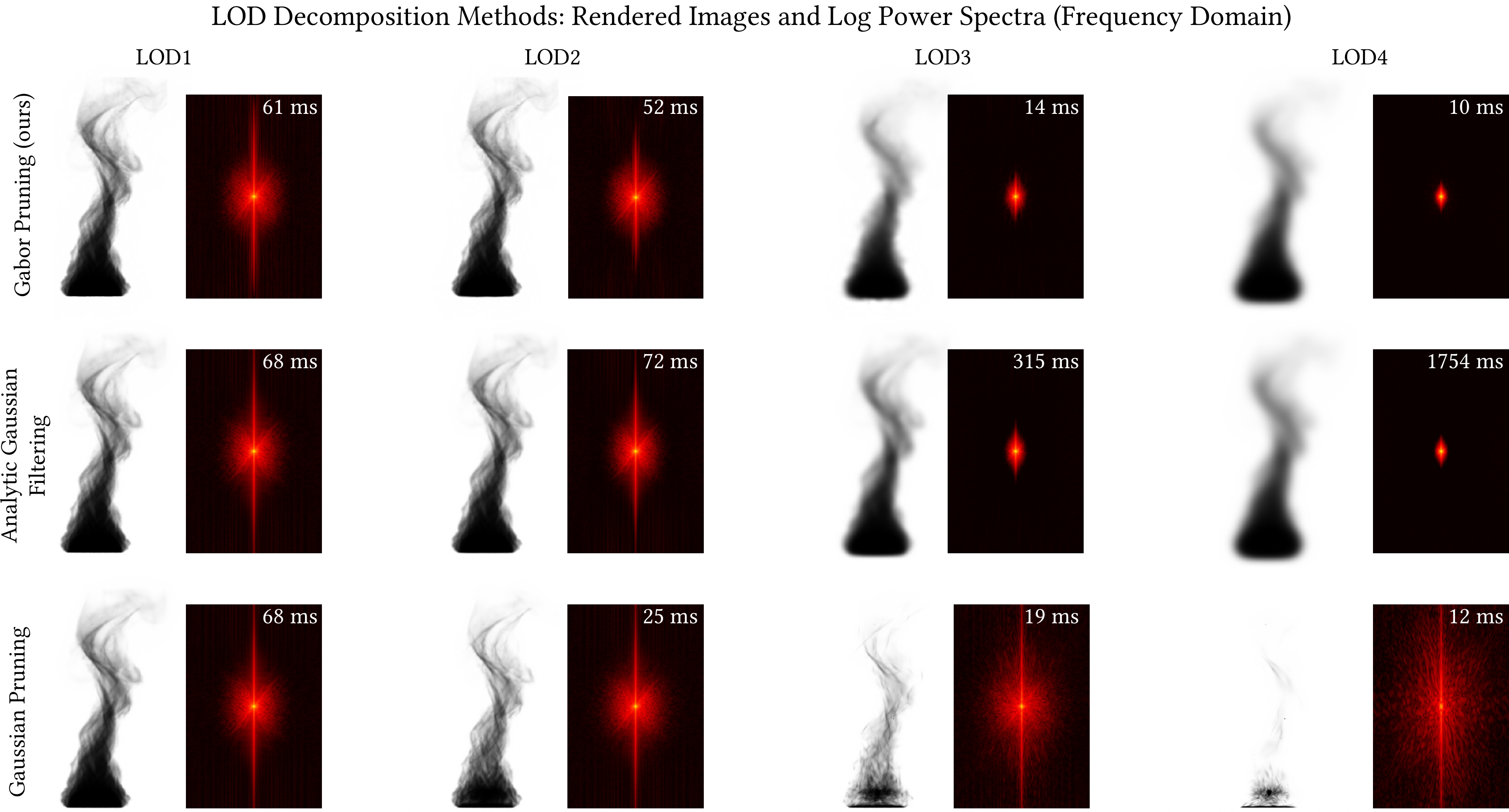}
    \caption{Comparison of different LOD strategies for primitive-based volumetric representations. In each cell, we show a purely absorbing medium's transmittance integral and its corresponding power spectra of the Fourier Transform. Our method closely replicates the effect of analytically filtering the Gaussian-based volume. However, filtering Gaussians analytically produces mixtures of equal complexity but with larger overlap, severely impacting performance. In contrast, our method composes the asset as a mixture of Gabor kernels, achieving filtering through simple removal of parts of the asset at runtime. While explored in radiance field rendering literature, we also show that a simple heuristic Gaussian pruning fails to create LOD levels with a low-pass characteristic.}
    \label{fig:lod_comparisons}
    \Description{A 3x4 grid comparing LOD strategies showing volume renderings and power spectra for Gabor Pruning, Gaussian Analytic, and Gaussian Pruning methods at different detail levels.}
\end{figure*}

As discussed in Section~\ref{sec:intro}, Gaussian primitives are not directly compatible with practical LOD strategies due to their spectral properties. Their power spectrum spans all frequencies centered at the origin. We visualize this in Figure~\ref{fig:lod_comparisons}. In essence, analytically filtering Gaussians to create LOD levels creates a new asset that retains the same number of Gaussians with larger spatial support. Consequently, this method does not yield any performance improvements. Pruning Gaussians based on their size is not a viable solution, as even small Gaussians carry important low-frequency information. Removing them results in artifacts at the coarser levels. An even more costly option would be to refit the pre-filtered asset using a smaller set of Gaussians. This would require solving multiple expensive optimization problems and would have higher memory requirements. Furthermore, interpolation between levels could introduce artifacts if the levels are not refitted jointly.

Our first insight is that, ideally, an effective primitive volume LOD strategy should be constructed like a Laplacian pyramid. A single asset could be decomposed into different LODs by simply removing a subset of primitives, allowing regression in a single pass, and continuous filtering during rendering, at no extra memory cost. This requires a kernel that, in the frequency domain, solely exists within specific frequency bands, effectively modelling a residual between the current and previous levels. This is where harmonic-modulated kernels come into play. In particular, we propose using an adaptation of the Gabor kernel~\cite{lagae2009procedural}. We extend the kernel by making it anisotropic, deriving closed-form line integrals for arbitrary segments, and a novel optimization scheme to regress volumetric density fields as mixtures of the kernel. We will then introduce several new strategies to leverage this volumetric representation to further accelerate rendering time in practical scenarios.

\subsection{The 3D Anisotropic Gabor Kernel}
\label{sec:gabor_kernel}
Our 3D anisotropic Gabor kernel has the form
\begin{multline}
    \hspace*{-1em}
    \label{eq:gabor_kernel}
    g(x; \mu, \Sigma, \vec{\omega}) = \frac{1}{\sqrt{8\pi^3\left | \Sigma \right |}} e^{-\frac{1}{2}(x-\mu)^T\Sigma^{-1}(x-\mu)}\cos{\left ( \vec{\omega}^T(x-\mu)\right )}
\end{multline}%
with $S$ the diagonal scale matrix (principal axis standard deviations) and rotation matrix $R$ of the Gaussian. Together they form a factorization of the covariance matrix $\Sigma = RSS^TR^T$~\cite{kerbl20233Dgaussians}; and $\vec{\omega}$ being the orientation and magnitude of our planar wave (i.e., 2D wave in 3D) modulation. In practice, we model the modulation $\vec{\omega}$ using a scalar $\omega$ and fix the orientation to be diagonal as $\vec{\omega}=RS^{-1}(\omega, \omega, \omega)^T$, making it relative to its envelope. This reduces the parameter count and simplifies the optimization. Our Gabor formulation is slightly different from the Gabor kernels typically used in Gaussian Textures~\cite{galerne12,filtering2011Lagae}, as the peak frequency of the modulation is entangled with the Gaussian's size. This is to naturally reinforce that smaller Gabor kernels cover higher frequencies (otherwise, the modulation and the envelope can be completely disentangled). 
We derive the closed-form solution for its analytical Fourier Transform and include detailed steps in the Appendix~\ref{sec:app_gaborfourier}. Given the peak frequency of the modulation $f_0 = \|\vec{\omega}\|$, its final derivation is
\begin{multline}
    \mathcal{F}\{g(x)\}(k) =\\ \frac{1}{2} e^{-ik^T\mu} \left[e^{-\frac{1}{2}(k - f_0)^T\Sigma (k - f_0)} + e^{-\frac{1}{2}(k + f_0)^T\Sigma (k + f_0)} \right]
\end{multline}%
Since we can analytically compute the spectral content of each Gabor kernel, we can easily obtain a good approximation of the band-limited filtered volume by removing the undesired kernels (Figure~\ref{fig:lod_comparisons}).

\subsection{Gabor Fields}
\label{sec:hierarchical_rep}
Inspired by Laplacian and Steerable pyramid decompositions~\cite{Simoncelli1995}, we pose primitive volumes as the composition of different kernel mixtures, each modelling specific spectral bands. If separated into discrete band-passes or levels, they compose a volumetric Laplacian pyramid, which we can further separate in different orientations to compose a "relaxed" Steerable pyramid (relaxed, as in we do not ensure orthogonality, due to the unlimited theoretical support of Gabor kernels). In practice, we combine positive, zero-frequency Gabor kernels (i.e., Gaussians) for the base of our pyramid (the low-pass), and Gabor kernels ($f_0>0$) for the remaining frequencies (residual). Gabor kernels can have both positive and negative density values, which discards them as a valid density function on their own. An example of our "pyramid decomposition" for a regressed asset can be found in Figure~\ref{fig:teaser}. While it is easier to think of it as separated "layers", in practice, we have total control over the frequency content, and continuous decompositions are perfectly possible. 

\subsection{Efficient Physically-based Rendering with Gabor Kernels}
\label{sec:rendering_gabor}
A key component in primitive-based volume rendering is the ability to compute transmittance integrals analytically. We thus derive closed-form definite line integrals for the projected 3D anisotropic Gabor kernel along an arbitrary ray. We include step-by-step derivations in the Appendix~\ref{sec:gabor_line_integral}, we present here the final integral expressions. For an arbitrary ray $r = x_0 + \vec{v} \cdot t$, and transforming the anisotropic Gabor into its canonical (i.e., whitened) local space, the full-domain integral of the projected Gabor kernel $g_r$ along the ray is

\begin{equation}
\int_{-\infty}^{\infty}g_r(x; \mu, \Sigma, \vec{\omega}) \,dt = \mathcal{K}\,e^{-\frac{1}{2}\left(c - b^2 + \Omega^2\right)}\cos{\left(d - \Omega b\right)}
\end{equation}%
For the segment-definite integral with general bounds $[t_0, t_1]$:
\begin{align}
\label{eq:definite_gabor_integral}
    \int_{t_0}^{t_1} g(\mathbf{r}, \omega, \Sigma, \mu)\,dt
    &= \frac{\mathcal{K}}{2} \,
    e^{-\frac{1}{2}\left(c - b^2 + \Omega^2\right)}
    \\
    &\quad \times \operatorname{Re}\!\left\{ e^{i\left(d - \Omega b\right)}\left[\operatorname{erf}\left(\frac{t_1 + b - i\Omega}{\sqrt{2}}\right) - \operatorname{erf}\left(\frac{t_0 + b - i\Omega}{\sqrt{2}}\right)\right] \right\}
    \nonumber
\end{align}%
with

\begin{equation}
    \begin{aligned}
    b &= \mathbf{p}_W \cdot \vec{v}_W, \quad 
    c = \mathbf{p}_W \cdot \mathbf{p}_W  \\[6pt]
    \Omega & = \vec{k}_W \cdot \vec{v}_W, \quad 
    d = \vec{k}_W \cdot \mathbf{p}_W \quad
    \mathcal{K} = \frac{1}{2\pi\sqrt{|\Sigma|} \cdot \|W\cdot\vec{v}\|}
    \end{aligned}
\end{equation}%
where we define the whitened ray parameters and frequency vector using ZCA whitening matrix $W$ as:

\begin{equation}
\begin{aligned}
W = R^T{}S^{-1}, 
\quad
\mathbf{p}_W = W(x_0 - \mu), 
\quad
\vec{v}_W = \frac{W\cdot\vec{v}}{\|W\cdot\vec{v}\|}
\quad
\vec{k}_W = W\cdot \vec{\omega}
\end{aligned}
\end{equation}%
Here $x_0$ and $\vec{v}$ are the ray origin and direction, $\mu$ and $\Sigma$ are the Gabor kernel mean and covariance, respectively. The whitened direction $\tilde{\mathbf{w}}$ is normalized so that $\|\vec{v}_W\| = 1$. The factor $\|W\cdot\vec{v}\|^{-1}$ is the Jacobian that arises as we normalize $\vec{v}_W$ to unit length for simpler integration. The integration parameter $t$ in whitened-normalized space relates to world space by this factor. The term $(c - b^2)$ represents the squared perpendicular distance from the kernel center to the ray in canonical space. The complex error function $\operatorname{erf}(\cdot)$ (Faddeeva function) handles the finite integration bounds.
\paragraph{Complex-valued Error Function} A key issue when integrating Gabor kernels analytically is the computation of complex-valued error functions. A number of algorithms can be found in the literature~\cite{gautschi1970efficient, poppe1990more, zaghloul2011}, but their accuracy, speed and ability to converge largely depend on the expected range of input arguments. 
In order to avoid branching on render time among many different algorithms, we implemented a truncated Taylor expansion~\cite{poppe1990more} for an effective tradeoff between performance and accuracy, ensuring its convergence for our expected value range. We thus approximate the $\operatorname{erf}$ term in Equation~\ref{eq:definite_gabor_integral} as:

\begin{equation}
    Re\left [\operatorname{erf}(z) \right ] \approx \frac{2}{\sqrt{\pi}} \sum_{n=0}^{N-1} \frac{(-1)^n \operatorname{Re}(z^{2n+1})}{n!(2n+1)}
\end{equation}%
with 
\begin{equation}
    z = \frac{t + b - i \Omega}{\sqrt{2}}
\end{equation}

\paragraph{Implementation details.} We implement it in DrJIT with early exits based on the relative size of newer terms in the series with respect to the previous term; in practice, we exploit the symmetries in the pair of $\operatorname{erf}$ terms to compute them jointly in a single loop. We truncate the series at a maximum of 16 terms to avoid numerical instability and bound the cost of the computation. Since this is a relatively expensive operation, we further optimize performance by catching segment sizes below a manually selected threshold ($\approx10^{-4}$ in whitened, local space units) and directly approximating their integral through the midpoint rule. This reduces the computation to a single Gabor probability density function evaluation. 

\paragraph{Distance Sampling} For the inversion of the cumulative density function (to enable distance sampling in multiple scattering), we resort to a bisection solver, as proposed by Condor et al~\shortcite{Condor2024Gaussians}. In practice, segments are generally very small, so uniformly sampling the segment, without inversion, is a good and cheap approximation we can optionally employ. While individual Gabor kernels do have areas where density is negative, it becomes irrelevant if the total mixture (considering all overlapping primitives, including Gaussians) is positive at any given 3D point.

\paragraph{Adaptive Clamping}
Ray-intersection queries and overlap processing dominate rendering time in volumetric primitive rendering. Reducing the extent (spatial support) of primitives greatly benefits rendering speed; however, reducing below $3\sigma$ could create artifacts (e.g. clipped primitives with obvious boundaries). The specific properties of Gabor kernels allow us to benefit particularly from spatial clamping while keeping error low. We develop a small heuristic to clamp the extent of each of our particles adaptively.

We start from Equation~\ref{eq:definite_gabor_integral}. Considering a perpendicular ray (maximum contribution), our goal is to find at which eccentricity from the kernel's mean the contribution becomes insignificant. While Gabor kernels have infinite zero-crossings due to their harmonic modulation, we can approximate that in the worst-case scenario (with a phase shift producing a cosine term of value 1), their contribution will be given by the attenuated integral $\rho$ along the perpendicular ray to $\vec{\omega}$ at extent $E = c - b^2$. 
\begin{equation}
    \rho(E) = \alpha \cdot\mathcal{K} \cdot \exp\left(-\frac{1}{2}(E^2 + \Omega^2)\right),
\end{equation}%
where $\alpha$ is the learned opacity scaler (Section~\ref{sec:regression}). Setting $\rho(E) = \epsilon$ for a threshold $\epsilon$ (minimum contribution to consider) and solving for the extent:

\begin{equation}
\label{eq:adaptive_extent}
    E_{\text{adaptive}} = \sqrt{-2\ln\left(2\pi\cdot \epsilon{\sqrt{|\Sigma|}} \right) - \vec{k}_W^2},
\end{equation}
The $\vec{k}_W^2$ term captures the frequency-induced decay: thus, higher-frequency Gabors can benefit from smaller extents. For Gaussian primitives ($f_0 = 0$), this simplifies to:
\begin{equation}
    E_{\text{adaptive}} = \sqrt{-2\ln\left(2\pi\cdot \epsilon{\sqrt{|\Sigma|}} \right)},
\end{equation}

\subsection{Frequency and Orientation-aware Volume Rendering}
\label{sec:pyramid_rendering_details}
Interesting opportunities for acceleration arise when a volume can be decomposed spectrally and angularly. In certain orientations, Gabor kernels have line integral values close to zero. This allows us to simply avoid ray-tracing subsets of the mixture depending on the ray orientation, or importance sample sets of particles based on their frequency, density, and orientation. In order to explain the concept of orientation selectivity, we plot several Gabor kernels with different frequencies in the spatial domain and their respective power spectrum plots in Fourier space in Figure~\ref{fig:selectivity_spectral}. We also include an animated version of this plot in the Supplementary material. In essence, the projected integral and frequency of a Gabor kernel along a ray is highly dependent on its orientation with respect to the modulation plane. In the extreme case of integrating along the plane itself, its integral tends to zero. Higher frequencies (right in Figure~\ref{fig:selectivity_spectral}) exhibit higher orientation selectivity. We further include the integral of an exemplary 3D Gabor kernel along a plane for visualization at different orientations in Figure~\ref{fig:selectivity_kernel}.

\begin{figure}
    \centering
    \includegraphics[width=\linewidth]{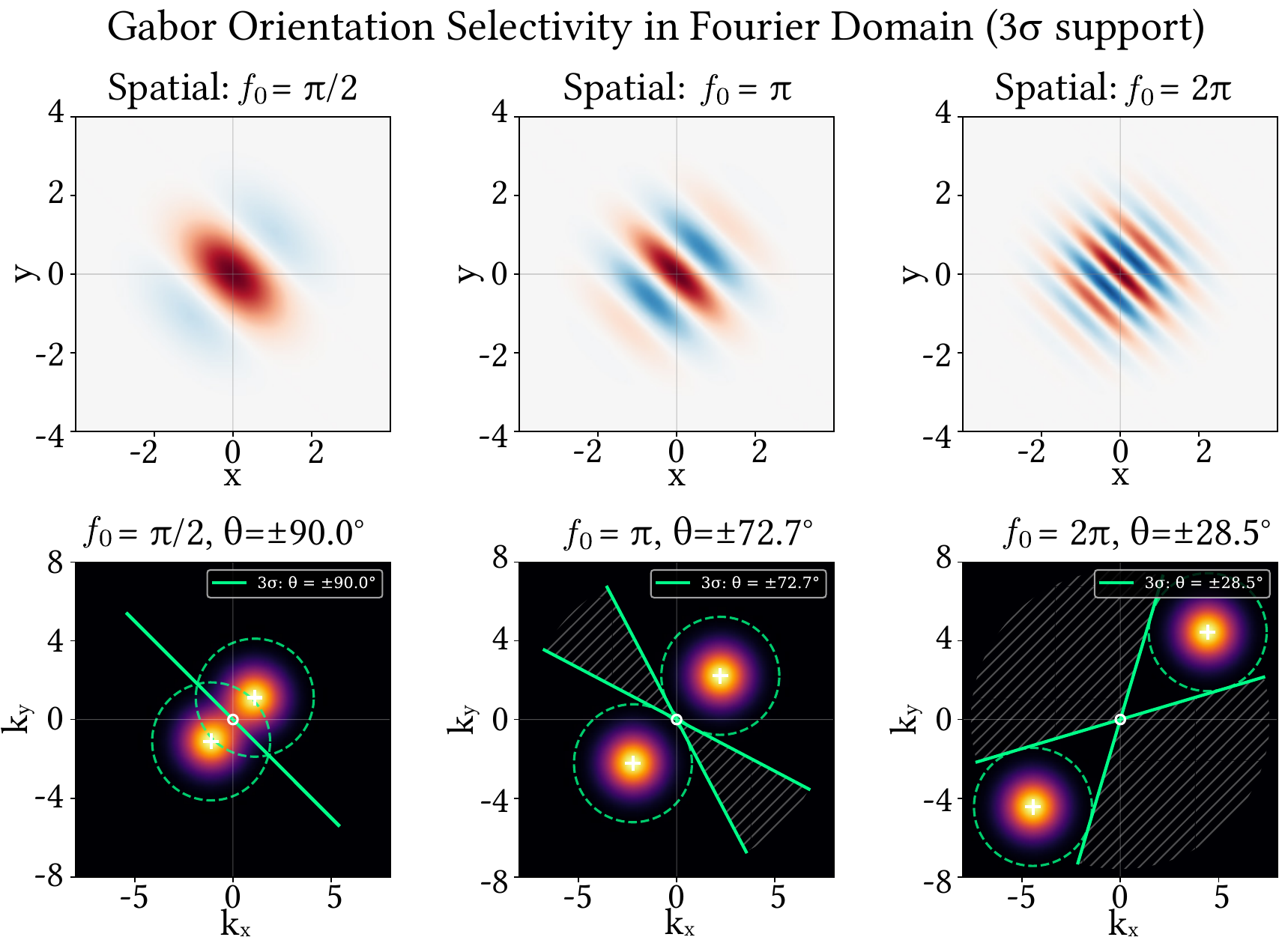}
    \caption{\textbf{Top:} 2D Gabor kernels in the spatial domain with different frequency magnitudes. \textbf{Bottom:} their respective power-spectrum representations. Gabor kernels exhibit directional attenuation, effectively modelling residuals in certain frequencies and orientations. As we clip their support to $3\sigma$, we can estimate at which orientations, relative to the integrating ray, their contribution decays enough to be considered  (shaded area). In our work, we leverage this insight to extract performance by masking non-contributing Gabor kernels before an intersection query is even traced.}
    \label{fig:selectivity_spectral}
    \Description{Two-row figure showing 2D Gabor kernels with increasing frequency in the spatial domain (top row) and their corresponding power-spectrum plots in Fourier space (bottom row), demonstrating directional attenuation.}
\end{figure}

\begin{figure}
    \centering
    \includegraphics[width=\linewidth]{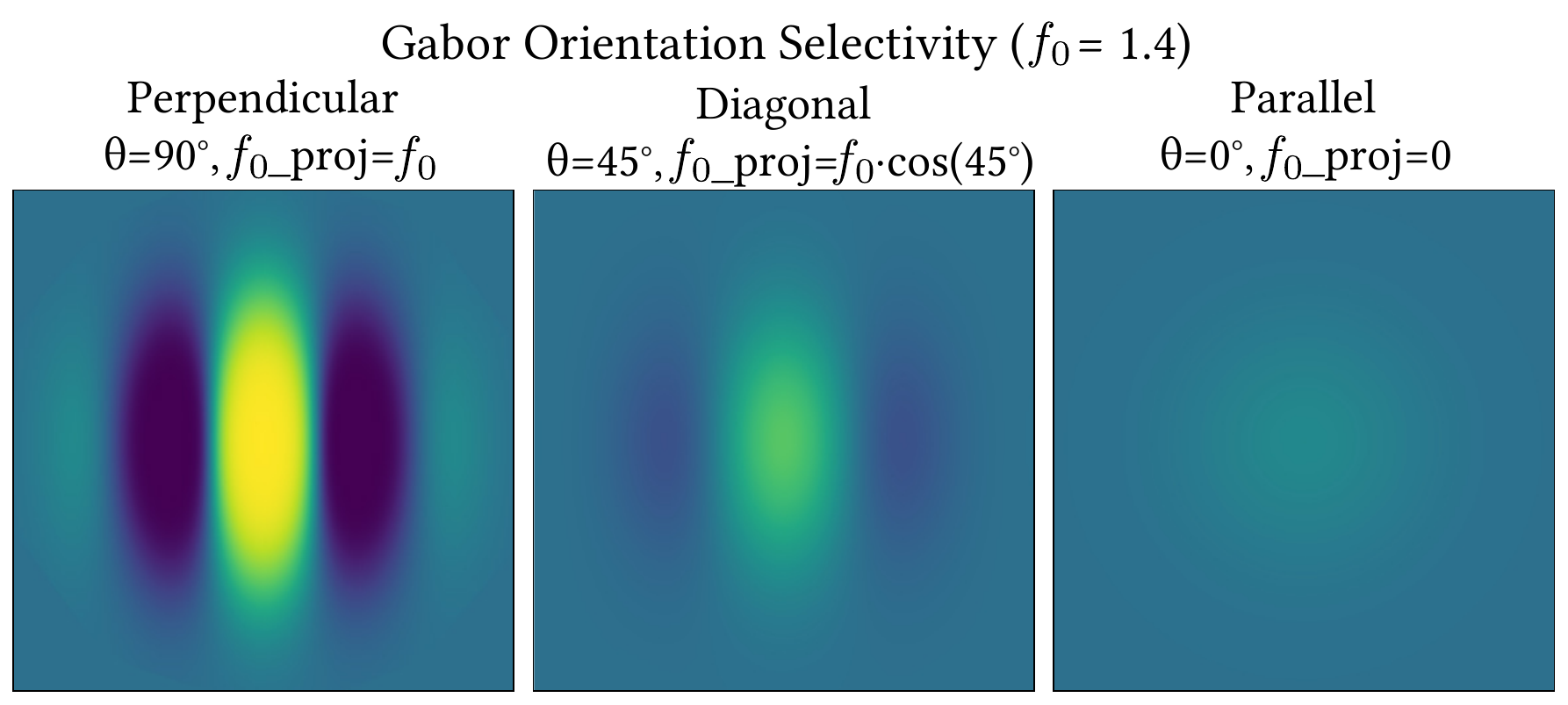}
    \caption{Visualization of the integral over a plane of a 3D Gabor kernel at different rotations around the Y axis. As we can see, projected frequency and integral magnitude both depend on orientation. In the limit, integrating along its modulation plane causes destructive interference that attenuates the} integral value significantly. Higher frequencies see their contributions decay even faster angularly, and to smaller integral magnitudes, as we show in our 2D example in Figure~\ref{fig:selectivity_spectral}.
    \label{fig:selectivity_kernel}
    \Description{A series of images showing planar integrals of a 3D Gabor kernel at different rotation angles around the Y axis, demonstrating how projection frequency and magnitude vary with orientation.}
\end{figure}

\subsubsection{Stochastic Laplacian Path Tracing}
Spectral selectivity enables a number of unorthodox acceleration strategies. Reducing mixture complexity and overlap for a given ray directly delivers substantial improvements in performance. In multiple scattering setups or complex scenes, a substantial amount of variance comes from the recursive nature of the integral itself. While obtaining exact transmittance integrals for every path segment reduces variance, much greater gains may be achieved by simply reducing traversal time across the volume and affording more samples. 

Inspired by recent work on stochastic texture filtering~\cite{fajardo2023stochastic} and procedural noise using Monte Carlo sampling~\cite{fajardo23procedural}, we develop a stochastic-analytical estimator: at render time, for each path segment, we stochastically decide which frequency bands of the asset to integrate. This essentially trades off increased variance (or bias, depending on the scenario) per sample with substantially faster traversal time across the volume (as in, reduced intersections in a simpler geometry structure). We propose a number of strategies in the Appendix (Table~\ref{tab:stochastic_sampling_strats}). In particular, we found the control-variate inspired techniques to be a good compromise between increased rendering performance and variance. This essentially means always including the low-pass (i.e., Gaussian band) within the set of primitives to intersect with, and additionally including another band (or multiple) of higher frequency Gabor kernels. This allows shifting variance or bias to lower or higher frequencies, as desired.

\subsubsection{Stochastic Orientation-selective Path Tracing}
Additionally to sampling from the frequency levels, we can exploit the angular selectivity of the volume to further accelerate rendering. For a 3D Gabor kernel $g(x;\mu,\Sigma,\vec{\omega})$ (Equation~\ref{eq:gabor_kernel}) integrated along a ray with direction $\vec{v}$, the contribution depends on the alignment angle $\theta$ between $\vec{v}$ and the frequency direction $\vec{\omega}$. Defining $a = |\cos\theta| \in [0,1]$ as the alignment factor: when the ray is perpendicular to the frequency direction ($a = 0$), the cosine modulation contributes uniformly along the ray. When parallel ($a = 1$), destructive interference reduces the contribution by $\mathcal{K}\exp(-f_{0}^2/2)$. We can then implement different strategies to reduce the number of kernels we trace against based on ray direction (Appendix Table~\ref{tab:steerable_sampling_strats}). In practice, we put the Gabor kernels in a pre-processing step into different bins based on the orientation and frequency of their modulations. During render time, for every new ray, and before we intersect with the scene, we compute its alignment with the representative bin orientations. Based on this alignment and the frequencies contained in the bin, we decide which primitives to mask out and which to include in the set to be "intersected". This choice can be deterministic (i.e. we simply prune low contributing orientations, paying with a small amount of error) or stochastic (i.e. we use importance sampling to sample proportionally from orientations contributing more to the given ray). For importance sampling, the sampling weight is essentially the dampened contribution by the projected frequency along the ray:
\begin{equation}
w(a) = e^{-\frac{f_{0}^2a^2}{2}} 
\end{equation}%
We showcase this in a Supplementary material video (rotating {\sc Bunny}). Both directional and Laplacian strategies can be used jointly, incrementing the performance boost.

\subsubsection{Analysis on Bias}
\label{sec:bias_analysis}
For individual segments, both proposed methods are unbiased. A renderer computing just ray transmittance integrals or tomographic reconstruction would be unbiased. However, this is not the case in multiple scattering settings, due to Jensen's inequality. Let $f$ be a convex function (in our case, an exponential) and $X$ a random variable. Jensen's inequality states:

\begin{equation}
    f\!\left(\mathbb{E}[X]\right)
    \;\le\;
    \mathbb{E}\!\left[f(X)\right].
\end{equation}%
resulting in a biased estimator of transmittance, despite our density integral estimator being unbiased. We can quantify the bias using a second-order Taylor expansion of $e^{-\hat{\tau}}$ around $\mathbb{E}[\hat{\tau}]$:

\begin{equation}
    \mathbb{E}[e^{-\hat{\tau}}] \approx e^{-\mathbb{E}[\hat{\tau}]} + \frac{1}{2} \text{Var}(\hat{\tau}) \cdot e^{-\mathbb{E}[\hat{\tau}]}
\end{equation}%
Therefore:
\begin{equation}
    \text{Bias} = \mathbb{E}[\hat{T}] - T_{\text{true}} \approx \frac{1}{2} \text{Var}(\hat{\tau}) \cdot T_{\text{true}} > 0
\end{equation}%
The bias is thus proportional to the variance of the optical depth estimator and the true transmittance (smaller $\tau$ means larger relative bias). Tuning the sampling strategy for specific needs and assets can reduce such bias (i.e. importance sampling based on the total energy of a pyramid level), or steer it in a perceptually-aware manner (by shifting error to specific spatial frequencies).

\subsection{Implementation Details}
\label{sec:implementation_details}
We implement our method in Mitsuba 3~\cite{jakob2022mitsuba3}, using a modified version of \emph{VPPT}~\cite{Condor2024Gaussians}. To maximize performance, we partition the assets at load time into a pyramid of frequencies and orientations, in order to leverage Optix ray visibility masking at the Top-level Acceleration Structure (TLAS). We extended Mitsuba 3 to support ray visibility masking, such that we can selectively decide which levels of the pyramid to integrate/interact with before even traversing the BVH. In practice, this means that each level has its own Geometry Acceleration Structure (GAS), which may be suboptimal but turns out to be more efficient than rejecting primitives after the intersection routine. 
More concretely, each pyramid level $l$ is assigned a visibility mask:
\begin{equation}
    \mathcal{V}_{l} = 2^l
\end{equation}%
The ray carries a mask $\mathcal{V}_{r}$ whose active bits are sampled according to some strategy (see Table~\ref{tab:steerable_sampling_strats}). We compute a bit-wise $\operatorname{AND}$ between ray and visibility masks during intersection tests. Only primitives with common active bits (e.g., $(\mathcal{V}_{r} \land \mathcal{V}_{l}) \neq 0$) are considered.
As an example, for a single ray, the rendering pipeline proceeds as follows:

\begin{enumerate}
    \item Sample a pyramid level based on the chosen strategy.
    \item Set the bits in $\mathcal{V}_{r}$ according to the sampling outcome.
    \item Perform ray-primitive intersection (filtered by mask) collecting overlapping primitives along the ray.
    
    \item Distance sampling (multiple-scattering, VPPT) or integrate the projected field directly (tomography).
    \item Reweight integrals based on sampling probabilities.
    \item Recurse ray (multiple scattering).
\end{enumerate}

In the tomography case, only after all samples are collected, we turn accumulated density into a transmittance estimate, assuming exponentially decaying media, as $T = \exp(-\tau)$; this makes it an unbiased estimator. While collecting the product of individual transmittances is mathematically equivalent, our density importance sampling strategy would make it a biased estimator (see Jensen's inequality above). In practice, primary path and secondary rays (i.e. next-event estimation, NEE) can also use different strategies: for NEE in particular, we can heavily skew the sampling towards the Gaussian level of the pyramid, which contains most of the energy in the volume, drastically reducing rendering time. We explore some of these in Section~\ref{sec:results} as well.

%% file: sections_paper/05_regression.tex
\section{Regressing Gabor Fields}
\label{sec:regression}
In order to transform discrete density fields (voxel grids) into Gabor Fields, we develop a novel regression scheme specifically tailored to our method. 

\subsection{Hierarchical Differentiable Rendering-Based Kernel Regression}
Similarly to stochastic sliced Wasserstein regression~\cite{kolouri2019generalized}, Condor et al~\shortcite{Condor2024Gaussians} proposed a regression scheme based on image-based differentiable rendering, optionally warm-started on a Gaussian mixture obtained through expectation-maximization.
While their approach is effective with simple and low-resolution assets, it does not scale well. This can be inferred by the limited complexity shown on high-resolution assets such as the Walt Disney Cloud\footnote{\url{https://disneyanimation.com/data-sets/}}. We attribute this to 1) the limited representation or modeling power of individual Gaussians and 2) scalability issues of an end-to-end regression scheme. Our proposed Gabor kernels improve on 1), and we leverage the strengths of these kernels by coupling them with a hierarchical regression procedure with tight controls over the frequencies modeled during regression.

\paragraph{Volume Filtering}
We create a Gaussian pyramid $G_l$ with a spatial standard deviation $\sigma_l$ that has the same perceived blur irrespective of grid resolution. Doing so simplifies the optimization of the lowest level, as the frequency range is fixed for all voxel grids. To achieve this, we work in a normalized continuous domain $\Omega = [0, 1]^3$. At level $l$, we define $\sigma_l = \sigma_02^l$, where $\sigma_0$ (in normalized units) controls the blur radius at the finest scale. A larger $\sigma_0$ corresponds to stronger blur at all levels. For a discrete voxel grid of resolution, $N_x \times N_y \times N_z$, let $N_{\max} = \max{(N_x, N_y, N_z)}$. We map normalized coordinates to voxel indices by $x = uN_{\max}$, $u \in \Omega$. The corresponding standard deviation in voxel units is
\begin{equation}
    \begin{aligned}
        \sigma^{\text{vox}}_l = \sigma_l N_{\max}.
    \end{aligned}\label{eq:vox_sigma}
\end{equation}

\paragraph{Hierarchical Regression}

To distill voxel density grids into a hierarchical Gabor field, we first build the Gaussian pyramid $\{G_l\}_{l=0,...,K}$ using \cref{eq:vox_sigma} with $\sigma_0 = \frac{1}{120\pi}$, which we determined empirically. The lowest level (typically 3 or 4) contains the lowest frequencies. All voxel grids are zero-padded to match $N_{\max}$ in all dimensions and placed in a $[-1, 1]^3$ cube to render reference images. We render tomographic reference images using ratio tracking \cite{novak14} with a high count of samples per pixel. We sample cameras from a hemisphere surrounding the $[-1, 1]^3$ cube.
We directly fit a Gaussian mixture to the lowest level. The remaining levels are jointly fit by a single Gabor mixture.

\paragraph{Initialization}
We derive analytic formulas to define the exact frequency band that our mixture has to model. The Gaussian kernels model the lowest frequency band, and their scale solely determines their spectral footprint. The smallest appropriate scale for each level $l$ is given by
\begin{equation}
    \begin{aligned}
        s_{\min}(l) = L \sigma_0 2^l,
    \end{aligned}
\end{equation}
where $L$ is the size of the continuous domain. The scale matching the signals' Nyquist frequency depends on the grid's resolution, and is given by
\begin{equation}
    \begin{aligned}
        s_\text{limit} = \frac{L \sqrt{2 \log \frac{1}{\epsilon}}}{\pi N_{\max}},
    \end{aligned}
\end{equation}
where $\epsilon$ is the effective cutoff, e.g., the magnitude response of the Gaussian filter at the cutoff frequency $|H(f_c)| = \epsilon$. We choose $\epsilon = 0.25$.
We sample the scales for the base layer using $s_{\text{high}} = s_{\min}(l_{\text{base}} + 1)$ and $s_{\text{low}} = s_{\min}(l_{\text{base}})$ by drawing from a Normal distribution:
\begin{equation}
    \begin{aligned}
        s_{\text{init}} \sim \left|\mathcal{N}\left(\mathbf{0}, \frac{s_{\text{high}} - s_{\text{low}}}{3}\right)\right| + s_{\text{low}}.
    \end{aligned}
\end{equation}%
For the Gabor kernels, we use $s_{\text{low}} = 1.5 \:s_\text{limit}\:\omega_\text{high}$ instead, thereby biasing the initialization to higher frequencies. We initialize the rotations randomly and $\omega_{\text{init}} \sim \operatorname{Unif}[0.7, 1.5]$. 
Centers are sampled from non-empty voxel locations to ensure good coverage, and opacities are drawn from $\alpha_{\text{init}}\mathcal{N}(1, \frac{1}{4})$ where $\alpha_\text{init}$ is the only hyperparameter adjusted depending on the reference voxel grid's density.

\paragraph{Optimization Procedure}
We follow the image-based differentiable rendering regression scheme introduced by \citeauthor{Condor2024Gaussians}~\shortcite{Condor2024Gaussians}. Since Gabor kernels are not valid probability distributions, we can not warm start using expectation-maximization. Inspired by \citeauthor{kheradmand20243d}~\shortcite{kheradmand20243d}, we define our optimization procedure as a stochastic sampling process from the underlying density distribution. This enables stronger exploration during early optimization and defines strategies to relocate unused primitives throughout the optimization. They show that optimization steps can be converted to Stochastic Gradient Langevin Dynamics by injecting noise into the gradients. The gradient update for a single Gaussian or Gabor kernel $\mathbf{g}$ becomes
\begin{equation}
    \begin{aligned}
        \mathbf{g} \leftarrow \mathbf{g} - \lambda_\text{lr} \cdot \nabla_{\mathbf{g}} \mathbb{E}_{\textbf{I}\sim\mathcal{I}}[\mathcal{L}(\mathbf{g}; \mathbf{I})] + \lambda_\text{noise} \cdot \epsilon,
    \end{aligned}
\end{equation}
where $\epsilon$ is the gradient noise facilitating exploration, $\lambda_\text{noise}$ controls the noise magnitude, $\lambda_\text{lr}$ is the learning rate, and $\textbf{I}$ is an image sampled from the training dataset $\mathcal{I}$. In practice, we compute the gradients over 32 renderings and use the Adam \cite{Kingma2014AdamAM}  optimizer update for  $\nabla_{\mathbf{g}} \mathbb{E}_{\textbf{I}\sim\mathcal{I}}[\mathcal{L}(\mathbf{g}; \mathbf{I})]$. We follow Kheradmand et al. \shortcite{kheradmand20243d} and only add gradient noise for the centers $\mu$.
The noise term is given by
\begin{equation}
    \begin{aligned}
        \epsilon_\mu = \lambda_\text{lr} \cdot \text{sigmoid}\left(-100(t - \alpha)\right) \cdot \Sigma\eta, \quad \eta \sim \mathcal{N}(\mathbf{0}, \mathbf{1}),
    \end{aligned}
\end{equation}
where $\alpha$, $\Sigma$ is the primitive's weight and covariance, $t$ is a hyperparameter and set to $1e-6$, which we later use as a threshold to determine primitives to be resampled through the MCMC optimization framework. The $\text{sigmoid}$ term reduces noise around the threshold to ensure a smooth decay for subsequent resampling. Essentially, we inject anisotropic noise identical to the primitive's Gaussian envelope.

\paragraph{Resampling Primitives}
For a stable optimization, it is crucial to ensure that the resampled primitives blend well into the current asset. The underlying idea is to teleport "dead" primitives (weight $\alpha_i < 1e-6$) to primitives that are alive and update their parameters to minimize visual change. The parameter update of \citeauthor{kheradmand20243d} was developed for rasterization; therefore we re-derive it for analytic ray tracing. All dead primitives will be teleported to a live primitive which we sample with probabilities proportional to their integral. The 3D integral of a Gabor kernel is given by
\begin{equation}
    \begin{aligned}
        \int g(x;\mu, \Sigma, \omega) = \frac{\alpha}{(2\pi)^{\frac{3}{2}}|\Sigma|^{\frac{1}{2}}e^{-\frac{3}{2}\omega^2}}.
    \end{aligned}
\end{equation}
Consider primitives $g_{1, ..., N-1}$ being teleported to the alive primitive $g_N$. The new parameters are given by:
\begin{equation}
    \begin{aligned}
        \mu^\text{new}_{1, ..., N} =  \mu^\text{old}_N, \quad \alpha^\text{new}_{1, ..., N} = \frac{\alpha_N^\text{old}}{N}\\
        \Sigma^\text{new}_{1, ..., N} =  \Sigma^\text{old}_N, \quad \omega^\text{new}_{1, ..., N} =  \omega^\text{old}_N,
    \end{aligned}
\end{equation}
This update keeps the density field of the mixture almost constant, except for the negligible loss of density at the location of the previously "dead" primitives. To further facilitate exploration, we keep the momentum statistics of Adam for the revived primitives but reset them for the sampled live primitives to discourage big movements. 

\paragraph{Loss Definition}
The main component of our loss is a mix of SSIM \cite{ssim} and L1. We further add regularization terms to favor smaller scales to improve render times and a weight decay, facilitating compact assets such that unused primitives can be resampled or later pruned. Opposed to Gaussian kernels, constraining the frequency content of anisotropic Gabor kernels is substantially more difficult, as constraining parameters independently will not suffice. As seen in Figure~\ref{fig:selectivity_spectral}, depending on their orientation and peak frequency of their harmonic modulation $\omega$, their spectral footprint changes drastically. To avoid aliasing, we derive a regularization term as a soft constraint. The angular cutoff frequency of each layer is given by
\begin{equation}
    \begin{aligned}
        \phi^{\text{cutoff}}(l) = \frac{\sqrt{2\log{\frac{1}{\epsilon}}}}{\sigma_0L 2^l},
    \end{aligned}\label{eq:layer_cutoff}
\end{equation}
and the angular Nyquist frequency yields the upper bound.

\begin{equation}
    \begin{aligned}
        \phi_{\text{limit}} = \frac{\pi N_{\max}}{L}
    \end{aligned}\label{eq:angular_nyquist}
\end{equation}
We propose
\begin{equation}
    \begin{aligned}
        \mathcal{L}_\text{freq} = \sum_i \text{sigmoid}\left(a (f_{0,i} -\phi_\text{limit})\right),
    \end{aligned}
\end{equation}
where $f_{0,i}$ is the peak frequency of a kernel and $a=10$ is a hyperparameter creating a steep transition, effectively activating the regularization term only when the peak frequency comes close to aliasing. The final loss term is given by
\begin{equation}
    \begin{aligned}
        \mathcal{L} &= 0.80 \cdot \mathcal{L}_\text{L1} + 0.20 \cdot \mathcal{L}_\text{SSIM} + \mathcal{L}_\text{freq} \\ &+ \lambda_\alpha \cdot \sum_i |\alpha_i| + \lambda_s \cdot \sum_{i}|s_{i}|_1 
    \end{aligned}
\end{equation}

\paragraph{Implementation Details}
We optimize our mixture models for 300 iterations using a cosine learning rate scheduler. We resample "dead" primitives every 30 iterations and have an exponential decay on the gradient noise level $\lambda_\text{noise}^\text{it} = \lambda_\text{noise}^0 \cdot \gamma^\text{it}$, where $\gamma = 0.65$. We include our final hyperparameters in the Appendix in Table~\ref{tab:hyperparameters}. We further ablated the influence of our components in an Ablation study in Table~\ref{tab:ablation_study}. Optimizing the low-frequency layer usually takes 2-8 minutes and the full volume between 0.1-2h (RTX 4090), depending on asset size and the number of primitives. 

%% file: sections_paper/06_results.tex
\section{Results}
\label{sec:results}
In this section, we compare our proposed modeling approach in terms of storage, regression quality, and rendering speed against previous works on classic voxel grid distributions and Gaussian primitive volumes~\cite{Condor2024Gaussians}. We also showcase novel rendering approaches that exploit the orientation- and frequency-selectivity of our decomposition to improve rendering performance and enable new applications. 

\subsection{Regression Quality and Compression}
\label{sec:results_regression}

\begin{figure}
    \centering
    \includegraphics[width=1.\linewidth]{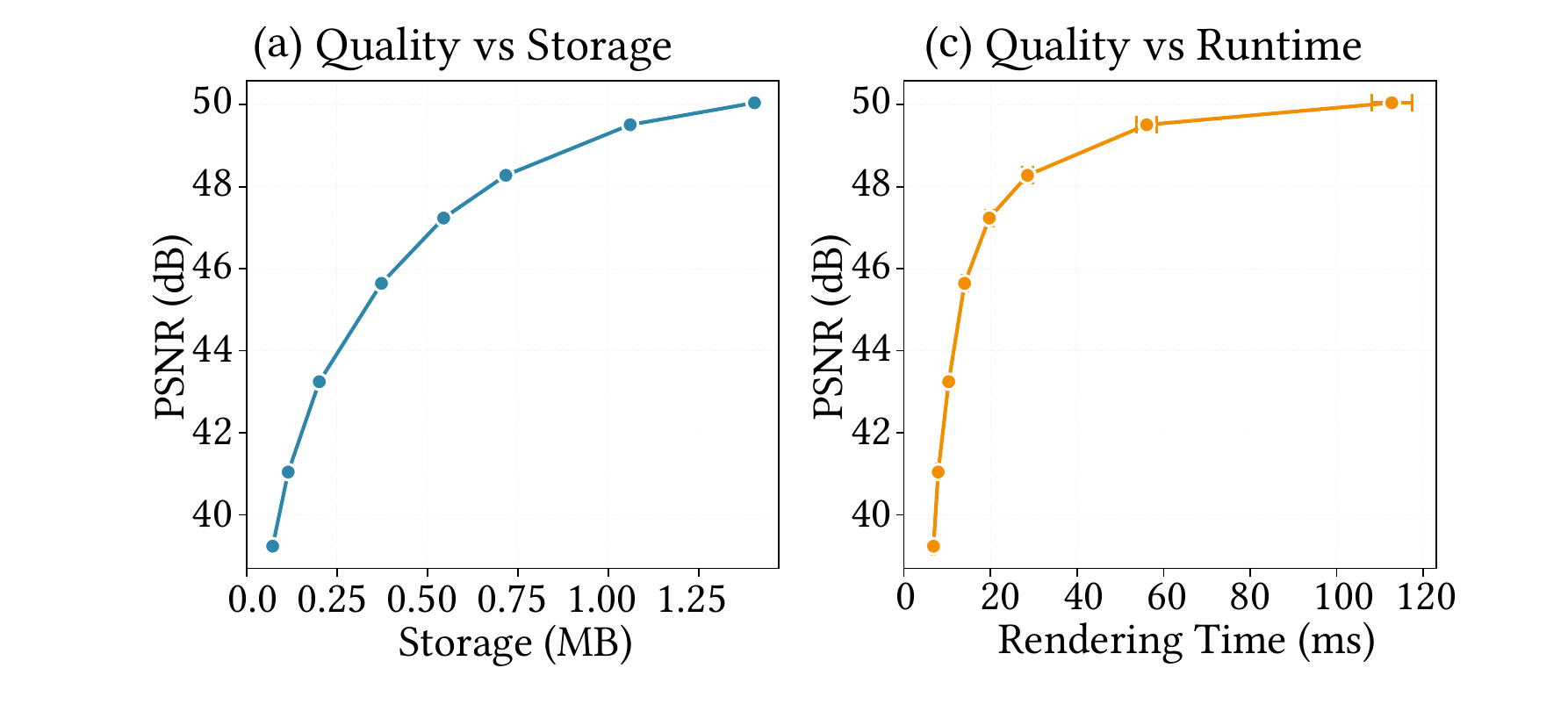}
    \caption{Scalability evaluation of Gabor-based representations on the  {\sc Tornado} dataset. The primitive count ranged from 1,024 to 32,768, with an additional 768 base Gaussian primitives. PSNR was measured with respect to the original voxel grid, and runtime was averaged over 50 runs. Storage costs assume 32-bit floating-point precision without compression. Measured on an RTX 4090.
    }
    \label{fig:scalability}
    \Description{Three line plots showing PSNR vs primitive count, render time vs primitive count, and storage vs primitive count for Gabor and Gaussian representations on the Tornado dataset.}
\end{figure}

\input{tables/ablation_table_simple}
\input{tables/comparison_table}
\input{tables/compression}
In Table~\ref{tab:gabor_vs_gaussian}, we compare Gabor Fields against Gaussian Primitive volumes~\cite{Condor2024Gaussians}, regressed with our own pipeline for a fair comparison, as the code for the full optimization process has not been released by the authors. We also include compression details in Table~\ref{tab:compression}, against compressed voxel grids, as well as Gaussian volumes. For the same number of primitives, Gaussian kernels are slightly more compact, as we include an extra parameter per particle ($\omega$). However, we also achieve superior quality, which becomes evident in the higher quality of details in Figure~\ref{fig:path_tracing}, where we showcase a variety of different assets. In Figure~\ref{fig:scalability}, we study how quality scales with the number of primitives for both Gabor and Gaussian assets. Further, we fit all individual frames of the animated {\sc{}Tornado} asset\footnote{\url{https://jangafx.com/software/embergen/download/free-vdb-animations}}, without hand-tuning of optimization parameters for individual volumes, to demonstrate the robustness and quality of our optimization. A rendered animation of the {\sc{}Tornado} asset is available in the Supplementary material. Additionally, we include an ablation to validate our choices in our regression method in Table~\ref{tab:ablation_study}. In general, Gabor Fields retain the compactness and quality of Gaussian Primitive Volumes, while providing fine-grained spectral control and a vast array of acceleration opportunities.

\subsection{Rendering Performance}
\label{sec:results_performance}

\subsubsection{Stochastic Laplacian and Orientation-selective rendering}
\begin{figure}
    \centering
    \includegraphics[width=\linewidth]{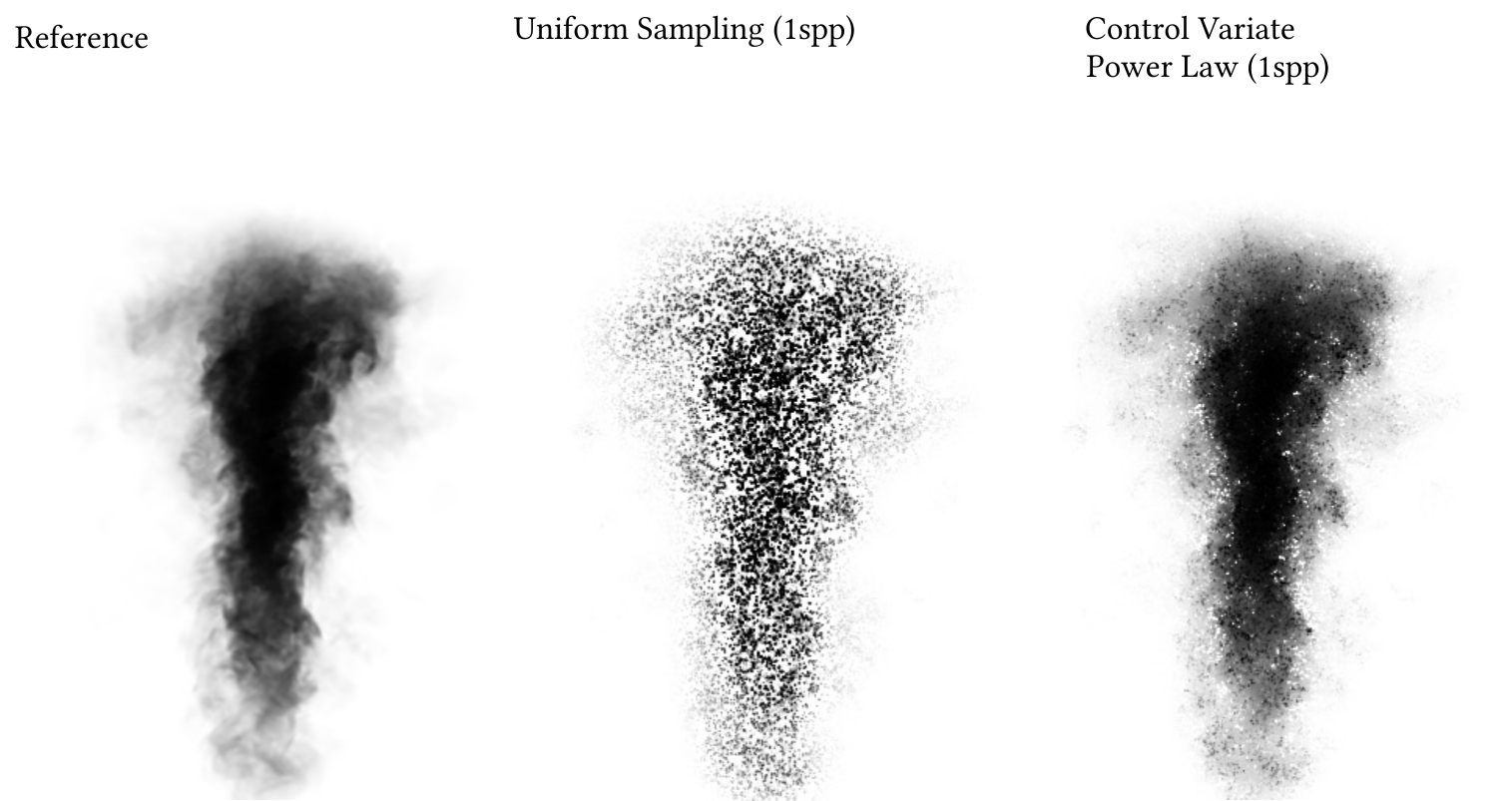}
    \caption{Some renders with our stochastic-analytical rendering methods ({\sc Tornado}).}
    \label{fig:stochastic_tomography}
    \Description{Multiple renderings of a tornado volumetric model using different stochastic-analytical rendering techniques, showing varying quality-performance tradeoffs.}
\end{figure}

\begin{figure}
    \centering
    \includegraphics[width=\linewidth]{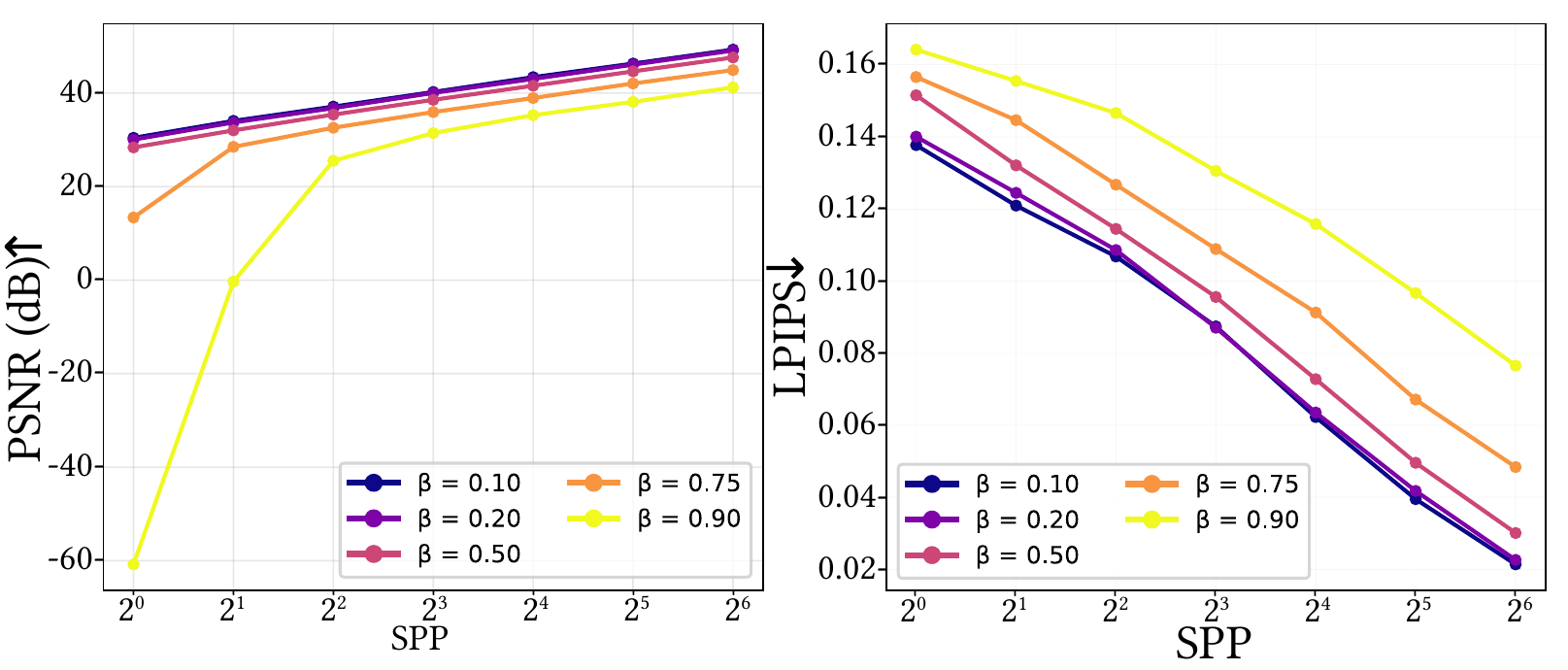}
    \caption{We ablate the choice of curve shape $\beta$ in our Control Variate Power Law strategy, for tomographic reconstruction in the {\sc Tornado} asset.}
    \label{fig:beta_ablation}
    \Description{A plot showing PSNR versus rendering time for different beta parameter values in the Control Variate Power Law strategy applied to the Tornado asset.}
\end{figure}

We showcase several configurations for stochastic Laplacian rendering in Figure~\ref{fig:stochastic_tomography}. We include an analysis on previously reported strategies in Figure~\ref{fig:quality_vs_time}. We also include a video showcasing orientation-selective strategies in the Supplementary material. While analytic rendering provides the best quality vs time tradeoff, stochastic methods can significantly reduce per-sample cost, which may be interesting in some applications. In tomographic rendering, these methods are unbiased. We also run an ablation with the {\sc Tornado} dataset trained with 33k particles on the choice of our parameter $\beta$ for the \emph{control variate power law} setup in Figure~\ref{fig:beta_ablation}. Substantial acceleration is possible for some increased variance. We also run an ablation with the {\sc Tornado} dataset trained with 33k particles on the choice of our parameter $\beta$ for the \emph{control variate power law} setup in Figure~\ref{fig:beta_ablation}. Substantial acceleration is possible for some increased variance.

\begin{figure}
    \centering
    \includegraphics[width=\linewidth]{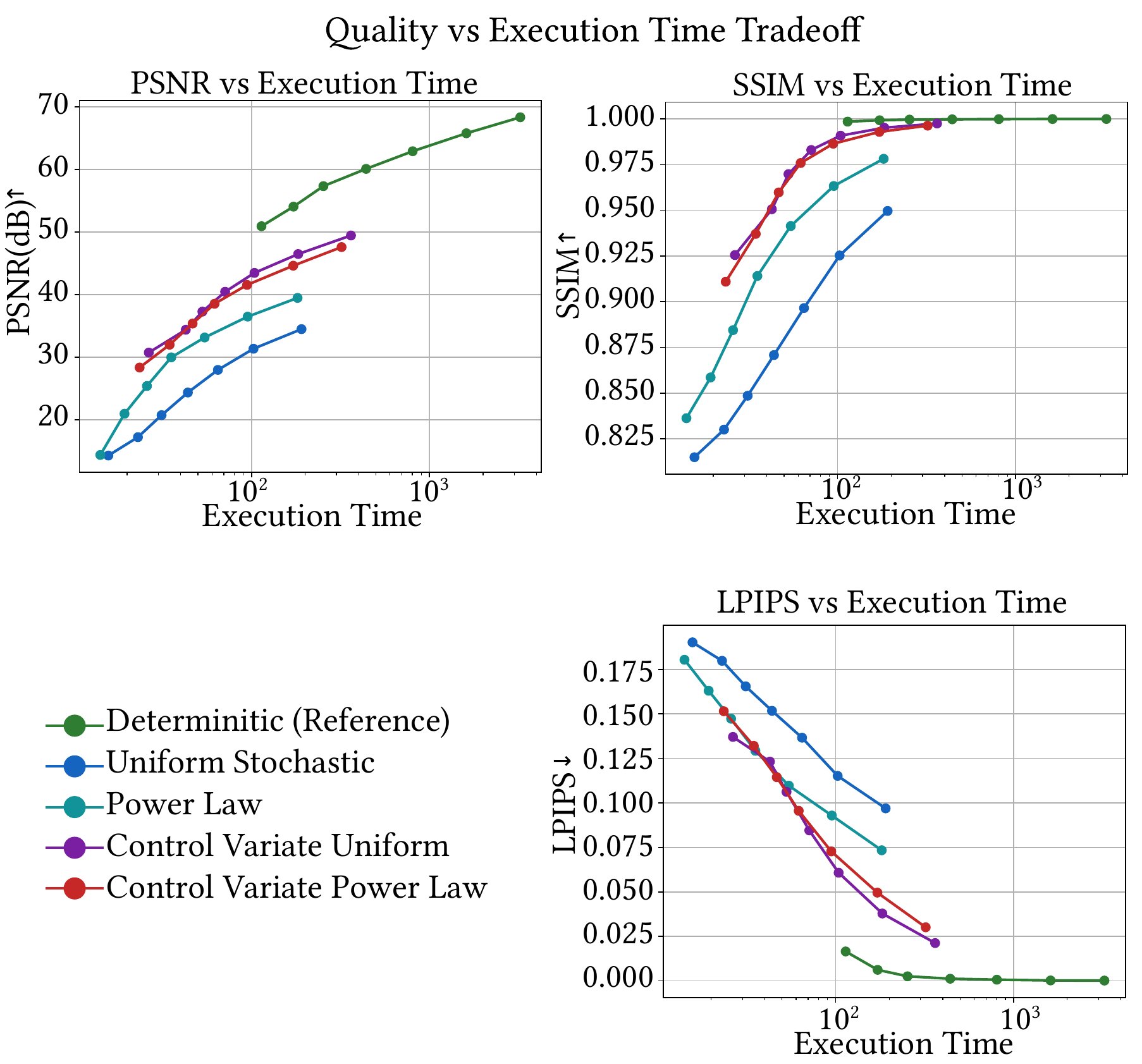}
    \caption{Quality vs rendering time with the different stochastic-analytical estimators (see Appendix Table~\ref{tab:stochastic_sampling_strats}). Starting from 1 spp, each dot represents powers of two extra samples per pixel (1, 2, 4, 8, etc).}
    \label{fig:quality_vs_time}
    \Description{A scatter plot showing PSNR quality versus rendering time for different stochastic-analytical estimators, with multiple lines representing different methods and dots showing sample counts.}
\end{figure}

\subsubsection{Path Tracing with Multiple Scattering}
\begin{figure*}
    \centering
    \includegraphics[width=\linewidth]{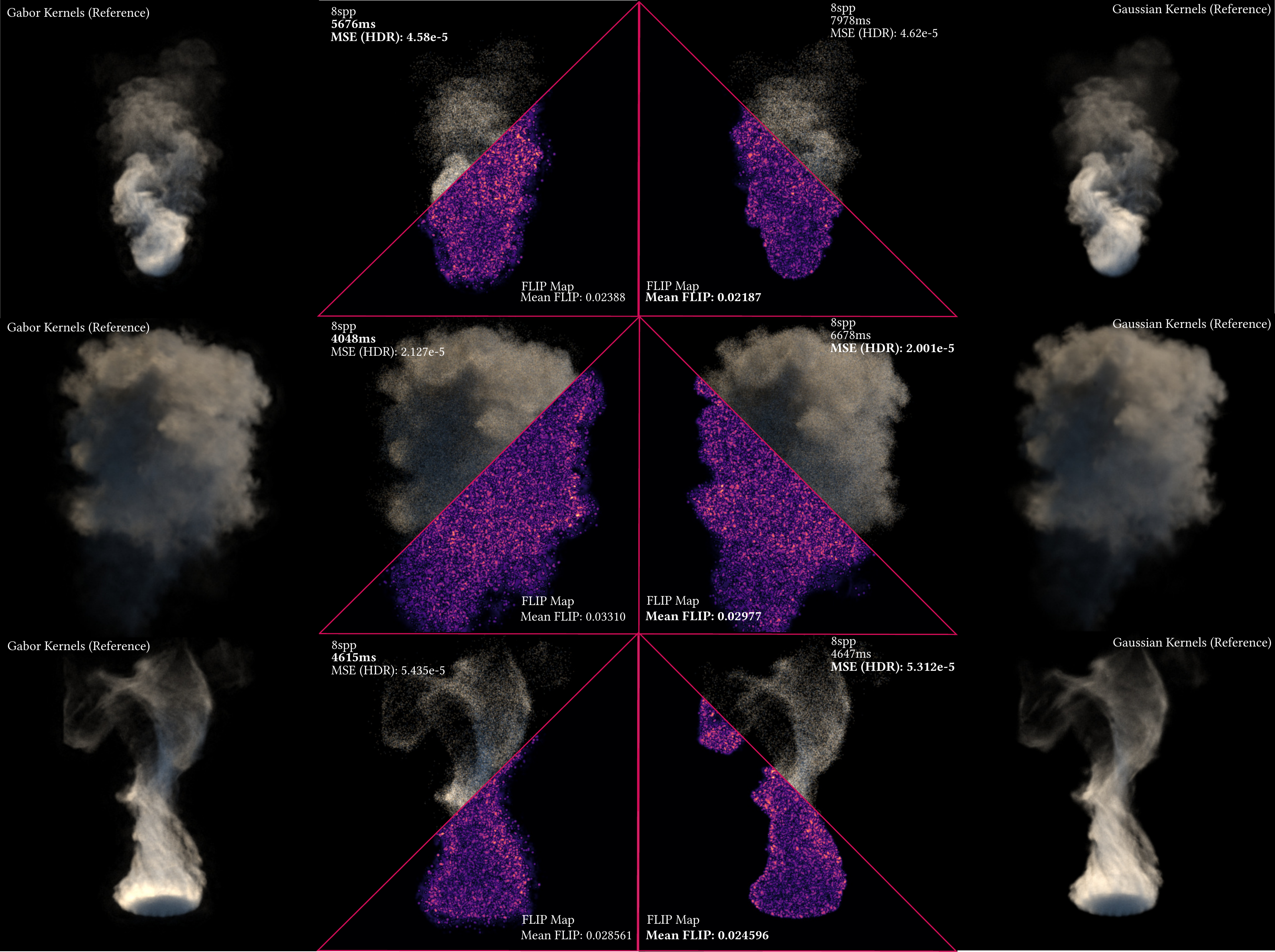}
    \caption{Comparison of Gabor Fields (Ours) vs Gaussian Primitive Volumes~\cite{Condor2024Gaussians}, for the assets, from top to bottom: {\sc{Smoke}}, {\sc{Explosion}} and {\sc{Dust Devil}}. For roughly equal variance and primitive count, Gabor Fields are not only faster, but higher quality, featuring higher frequency detail, as reported in Table~\ref{tab:gabor_vs_gaussian}. Times reported measured on an RTX 4090.}
    \label{fig:path_tracing}
    \Description{Side-by-side comparisons of multiple volumetric scenes rendered with Gabor Fields versus Gaussian Primitive Volumes, showing visual quality and rendering time differences for various datasets, including clouds and smoke.}
\end{figure*}

\begin{figure*}
    \includegraphics[width=\linewidth]{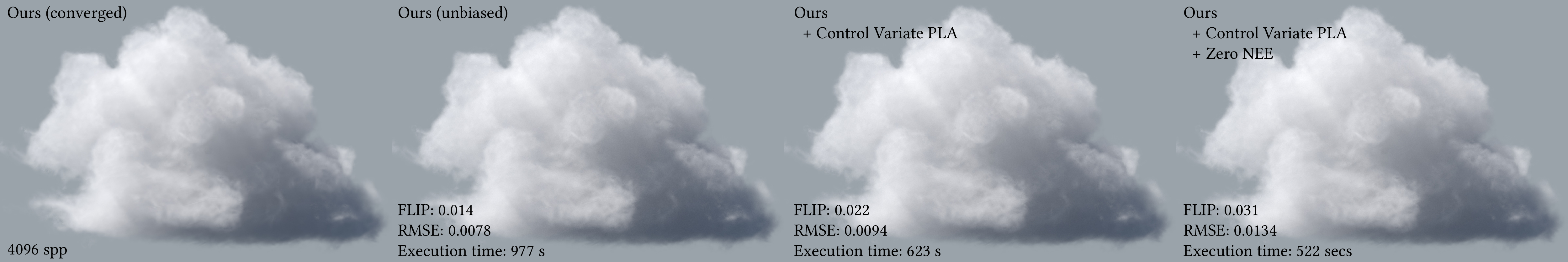}
    \caption{Showcase of different acceleration strategies enabled by our method with a {\sc{}WDAS Cloud} dataset with 42k Gabor Kernels, and 512 spp. Control Variate Power Law Accumulating (see Appendix Table~\ref{tab:stochastic_sampling_strats}) is used on primary paths, biasing slightly at high frequencies, given our choice of $\beta = 0.2$. Zero NEE indicates that for secondary shadow rays (next-event estimation), we only trace against the Gaussian level. This is a powerful strategy, as most of the energy in the density field is contained in the Gaussian level, with Gabors only modelling residual densities. This yields a substantial performance gain in exchange for some amount of error, which may be acceptable in some situations. Other alternatives, such as tracing against Gaussians only in the primary path after a number of recursions, are also explored in Supplementary materials. Timings reported on an RTX 6000. Asset is part of the Walt Disney Animation Studios \emph{cloud} dataset (CC-BY-SA 3.0)}
    \label{fig:ablation_biased}
    \Description{Comparison of cloud renderings showing different biased acceleration strategies with varying rendering times and quality, demonstrating tradeoffs between performance and accuracy for the Disney Cloud dataset.}
\end{figure*}

We compare rendering performance and quality against Gaussian Primitive Volumes~\cite{Condor2024Gaussians} in Figure~\ref{fig:path_tracing}. Additional examples of higher quality renders with our Gabor Field Path Tracer can be seen in Figures~\ref{fig:teaser},~\ref{fig:ablation_biased} and~\ref{fig:army_bunny}. For a fair comparison, we allow Gaussian volumes to benefit from the same acceleration strategies we describe for Gabor Fields in Section~\ref{sec:implementation_details}: small-segment integral approximation, adaptive clamping and whitened-space integration. Still, our Gabor Fields routinely outperform Gaussian Primitive Volumes in performance at roughly equal variance. This is still without any frequency or angular-based acceleration technique, rendering at the highest possible detail. On top of improved performance, we can observe higher quality detail, particularly in the higher frequency content (e.g., smoke plume on the first row); this is in line with the superior regression results we report in Table~\ref{tab:gabor_vs_gaussian}. At the same time, if LOD is required, our method will also outperform Gaussian volumes in terms of memory, as it can be naturally filtered right away, while Gaussian Primitives would need additional storage for e.g. a volumetric mipmap (Figure~\ref{fig:lod_comparisons}). 

\paragraph{Biased Acceleration strategies}
The amount of bias with spectral and orientation-selective acceleration techniques depends largely on the scene and strategy selected; we do a short showcase of an example scene in Figure~\ref{fig:ablation_biased}. As can be observed, substantial performance can be extracted for a small exactness or quality price, with full control over the spectral content of the error.

\subsection{Level of Detail}
A prime application of our method is providing seamless LOD. We showcase it in Figures~\ref{fig:teaser},~\ref{fig:army_bunny} and ~\ref{fig:lod_comparisons} and in the Supplementary materials, where we include videos of cameras zooming in and out of one of our assets, and a static visualization for comparison. Particularly, in Figure~\ref{fig:lod_comparisons}, we compare it against two baseline approaches for providing LoD in Gaussian-based volumes: analytical filtering of the Gaussians, and pruning-based bandlimiting. While the former provides perfect filtering, it does so at a substantial performance and memory cost. In the second approach, since Gaussian modelling high frequencies also model low frequencies (their power spectrum's mean is located at 0), pruning high-frequency Gaussians degrades quality. This can be seen more prominently in the Fourier power spectra of the three examples. In contrast, our work seamlessly supports LoD rendering; for any given desired frequency limit, we can select the minimal asset that matches the criteria by simply masking out part of the Gabor kernels. 
\begin{figure*}
    \centering
    \includegraphics[width=0.48\linewidth]{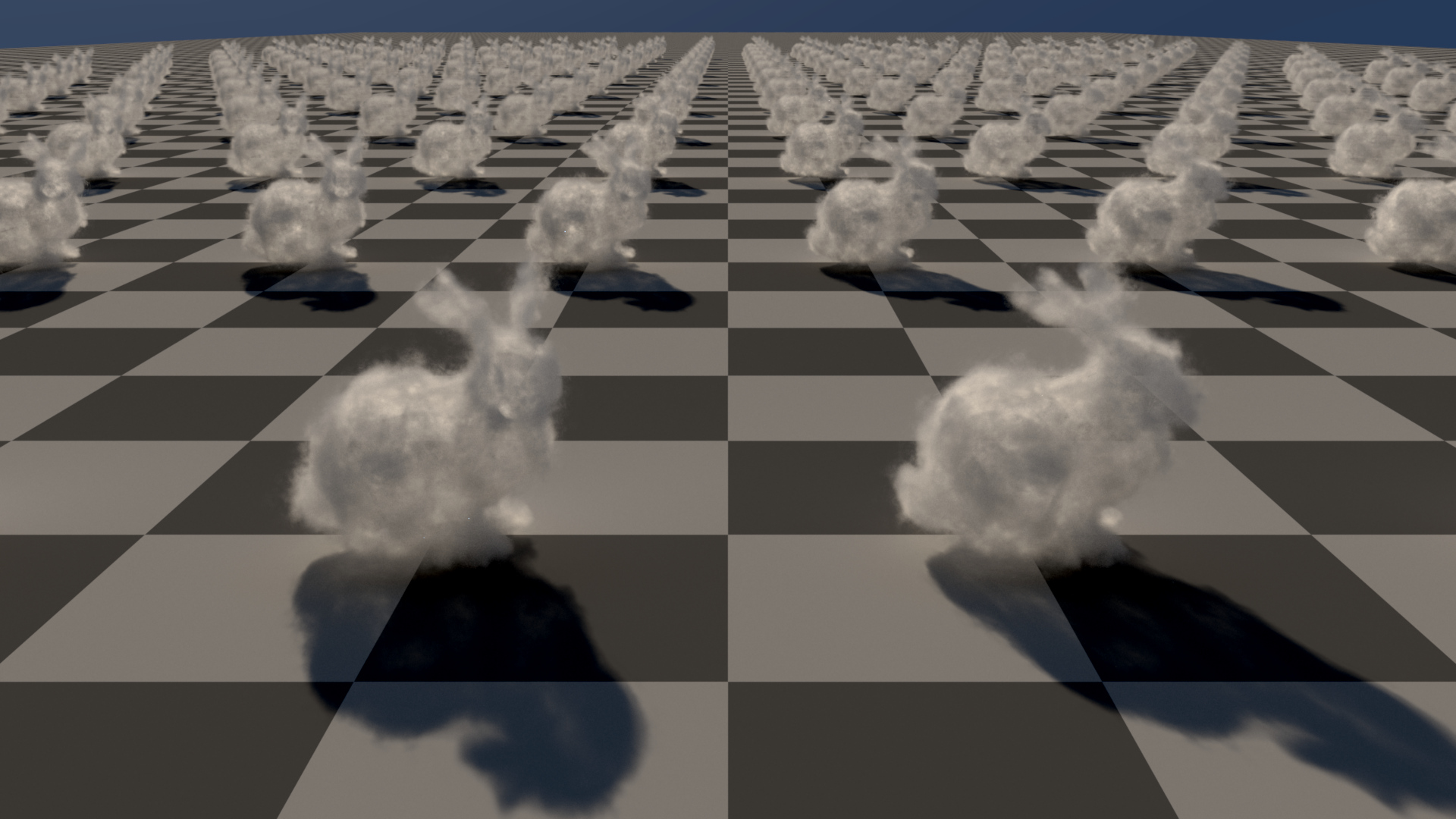}
    \includegraphics[width=0.48\linewidth]{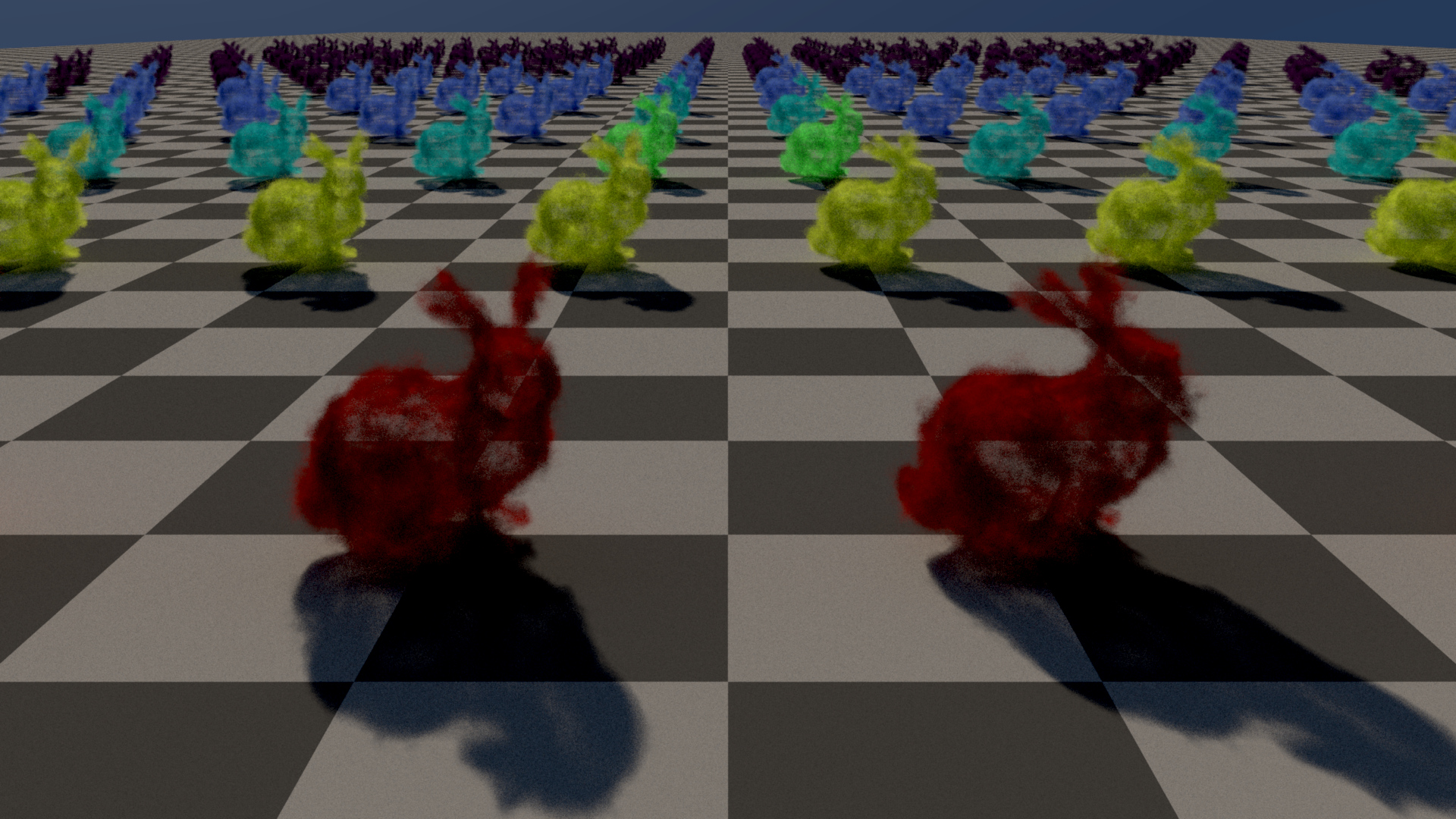}
    \caption{Stress-test of our LOD strategy using multiple scattering. Each {\sc{}Bunny} is composed of 33,280 Gabor kernels. We place 176 of them, which amounts to 5,857,280 primitives. Using our Laplacian decomposition, we can directly filter them on demand during render time, significantly accelerating rendering speed and eliminating aliasing. In the end, this frame featured an effective count of 759,164 primitives (x7.7 smaller). We visualize the chosen LOD levels on the right with colorized albedos.}
    \label{fig:army_bunny}
    \Description{Two images showing an army of 176 volumetric bunnies with LOD applied. The left shows the rendered result at 800 samples per pixel, and the right shows the same scene with color-coded LOD levels at 32 samples per pixel.}
\end{figure*}

%% file: tables/ablation_table_simple.tex
\begin{table}[t]
  \centering
  \small
  \caption{
    Ablation study on the {\sc Bunny} volume with 300$\times$300 resolution.
    PSNR $\Delta$ shown relative to the full method. Runtime was averaged over 50 runs on an RTX 4090.
  }
  \begin{tabular}{lcccccc}
    \toprule
    \textbf{Ablation} & \textbf{PSNR}$\uparrow$  & $\Delta$$\uparrow$  & \textbf{SSIM}$\uparrow$ & \textbf{LPIPS}$\downarrow$ & \textbf{Render}$\downarrow$ \\
    & \textbf{(dB)} & \textbf{(dB)} &  &  & \textbf{(ms)} \\
    \midrule
    \textbf{Full Method} & \textbf{48.86} & \textbf{+0.00} & \textbf{0.9996} & \textbf{0.0073} & \textbf{24.1} \\
    w/o Scale Regu. & 48.78 & -0.08 & 0.9996 & 0.0089  & 37.7 \\
    w/o Trainable $\omega$ & 48.66 & -0.20 & 0.9995 & 0.0078 & 27.3 \\
    w/o Frequency Regu. & 46.95 & -1.92 & 0.9994 & 0.0102  & 26.9 \\
    w/o Importance\\ \quad Sampling Centers & 46.36 & -2.50 & 0.9993 & 0.0170  & 38.0 \\
    w/o Opacity Regu. & 45.50 & -3.37 & 0.9991 & 0.0133  & 31.0 \\
    \bottomrule
  \end{tabular}

  \label{tab:ablation_study}
\end{table}

%% file: tables/comparison_table.tex
\begin{table*}[htbp]
  \centering
  \small
  \caption{Comparison of Gabor and Gaussian kernel performance on volumetric assets tomography regression. We report {\sc{WDAS\ Cloud}} at $\frac{1}{8}$ resolution.}
  \begin{tabular}{>{\raggedright\arraybackslash}p{0.16\linewidth}rrrrrrrrrrrr
  }
    \toprule
    \textbf{Asset} & \multicolumn{2}{c}{\textbf{PSNR $\uparrow$}} & \multicolumn{2}{c}{\textbf{SSIM $\uparrow$}} & \multicolumn{2}{c}{\textbf{L1 $\downarrow$}} & \multicolumn{2}{c}{\textbf{L2 $\downarrow$}} & \multicolumn{2}{c}{\textbf{LPIPS $\downarrow$}} & \multicolumn{2}{c}{\textbf{\#Primitives $\downarrow$}} \\
     & \textit{Gabor} & \textit{Gauss} & \textit{Gabor} & \textit{Gauss} & \textit{Gabor} & \textit{Gauss} & \textit{Gabor} & \textit{Gauss} & \textit{Gabor} & \textit{Gauss} & \textit{Gabor} & \textit{Gauss}\\
    \midrule
    \sc{Dust Devil} & \textbf{52.54} & 49.12 & \textbf{0.9997} & 0.9987 & \textbf{0.000868} & 0.001382 & \textbf{0.000006} & 0.000012 & \textbf{0.0068} & 0.0161  & 24,576 & {24,576} \\
    \sc{Explosion} & \textbf{53.12} & 51.47 & \textbf{0.9998} & 0.9997 & \textbf{0.001072} & 0.001335 & \textbf{0.000005} & 0.000007 & \textbf{0.0030} & 0.0067  & 24,576 & {24,576} \\
    \sc{Fire} & \textbf{51.13} & 49.22 & \textbf{0.9996} & 0.9988 & \textbf{0.001218} & 0.001692 & \textbf{0.000008} & 0.000012 & \textbf{0.0145} & 0.0226 &  24,576 & {24,576} \\
    \sc{Smoke} & \textbf{50.08} & 47.20 & \textbf{0.9996} & 0.9982 & \textbf{0.001190} & 0.001999 & \textbf{0.000010} & 0.000019 & \textbf{0.0087} & 0.0187 & 16,384 & {16,384} \\
    \sc{WDAS Cloud} & \textbf{47.15} & 46.92 & 0.9987 & \textbf{0.9988} & 0.002045 & \textbf{0.002044} & \textbf{0.000019} & 0.000020 & 0.0424 & \textbf{0.0416} & 24,576 & 24,576 \\
    \sc{Dust} & \textbf{42.30} & 40.73 & \textbf{0.9985} & \textbf{0.9985} & \textbf{0.003774} & 0.004396 & \textbf{0.000059} & 0.000084 &  0.0930 & \textbf{0.0781} & 42,496 & 42,496 \\
                     
    \midrule
    \textbf{Average} & \textbf{49.39} & 47.44 & \textbf{0.9993} & 0.9988 & \textbf{0.001695} & 0.002141 & \textbf{0.000018} & 0.000026 & \textbf{0.0281} & 0.0306 & 26,197 & 26,197 \\
    \bottomrule
  \end{tabular}
  \label{tab:gabor_vs_gaussian}
\end{table*}

%% file: tables/compression.tex
\begin{table}[htbp]
  \centering
  \small
  \caption{Memory footprint analysis for Gabor and Gaussian mixtures with an equal number of primitives, optimized for maximum quality vs. memory footprint of a voxel grid. The superscript of {\sc{WDAS\ Cloud}} represents the downsampling factor used.}
  
  \begin{tabular}{lrrrrrr}
    \toprule
    \textbf{Asset} & \textbf{Voxels} & \textbf{\#Prim} & \textbf{Gabor} & \textbf{Gauss} & \textbf{Voxel Grid} \\
     &  &  & \textbf{(MB)} & \textbf{(MB)} & \textbf{(MB)} \\
    \midrule
    \sc{WDAS\ Cloud}$^{\frac{1}{1.4}}$ & 2.04 B & 46,749 & 1.95 & 1.78 & 7764.75 \\
    \sc{Bunny} & 144.6 M & 32,768 & 1.38  & 1.25 & 578.24 \\
    \sc{Tornado} & 78.7 M & 24,576 & 1.03  & 0.94 & 314.90 \\
    \sc{Dust} & 58.9 M & 42,496 & 1.78 & 1.62 & 225.00 \\
    \sc{Dust\ Devil} & 18.7 M & 24,576 & 1.03  & 0.94 & 71.16   \\
    \sc{WDAS\ Cloud}$^\frac{1}{8} $ & 13.0 M & 24,576 & 1.03 & 0.94 & 49.77   \\
    \sc{Explosion} & 12.6 M & 24,576 & 1.03  & 0.94 & 47.97   \\
    \sc{Fire} & 9.0 M & 24,576 & 1.03 & 0.94 & 34.20   \\
    \sc{Smoke} & 2.8 M & 16,384 & 0.69  & 0.62 & 10.53   \\
    \bottomrule
  \end{tabular}
  \label{tab:compression}
\end{table}

%% file: sections_paper/07_applications.tex
\section{Other Applications}
\label{sec:other_applications}
Our representation can be useful in other applications where frequency and/or orientation selectivity is critical. Here we demonstrate applications to foveated rendering, motion blur and procedural asset generation.

\subsection{Foveated Rendering}
\label{sec:foveated_app}
Similar to LOD control, our Gabor-based representation can be used for foveated rendering to reduce rendering costs when the observer's gaze location is provided \cite{mohanto-2022,guenter-2012,surace-2023}. A common approach is to reduce the rendering resolution in peripheral vision, which can be easily achieved by progressively discarding Gabor primitives as the distance from the gaze location (i.e., eccentricity) increases. This can be done on a per-ray basis at render time. We discuss our strategy with more detail in the Supplementary material.
Figure~\ref{fig:foveation} demonstrates an example of applying our strategy. Our method reduces both the primitive count via level masking and the integration cost by discarding individual primitives. The primary bottleneck is still BVH traversal. Our approach can be further combined with stochastic and directional rendering strategies (Section~\ref{sec:pyramid_rendering_details}) to reduce sample cost at the periphery. Furthermore, higher acceleration can be achieved with adaptive resolution. In our demonstration, we focus only on truncating the representation. Finally, given that real-time ray tracing suffers from limited sample budgets, combining tight control over spectral content with post-rendering filtering techniques, such as \cite{DLSS}, can yield better results. For example, accounting for denoising capabilities could enable more aggressive removal of the Gabor primitives~\cite{karpenko-2025}.

\begin{figure}
    \centering
    \includegraphics[width=\linewidth]{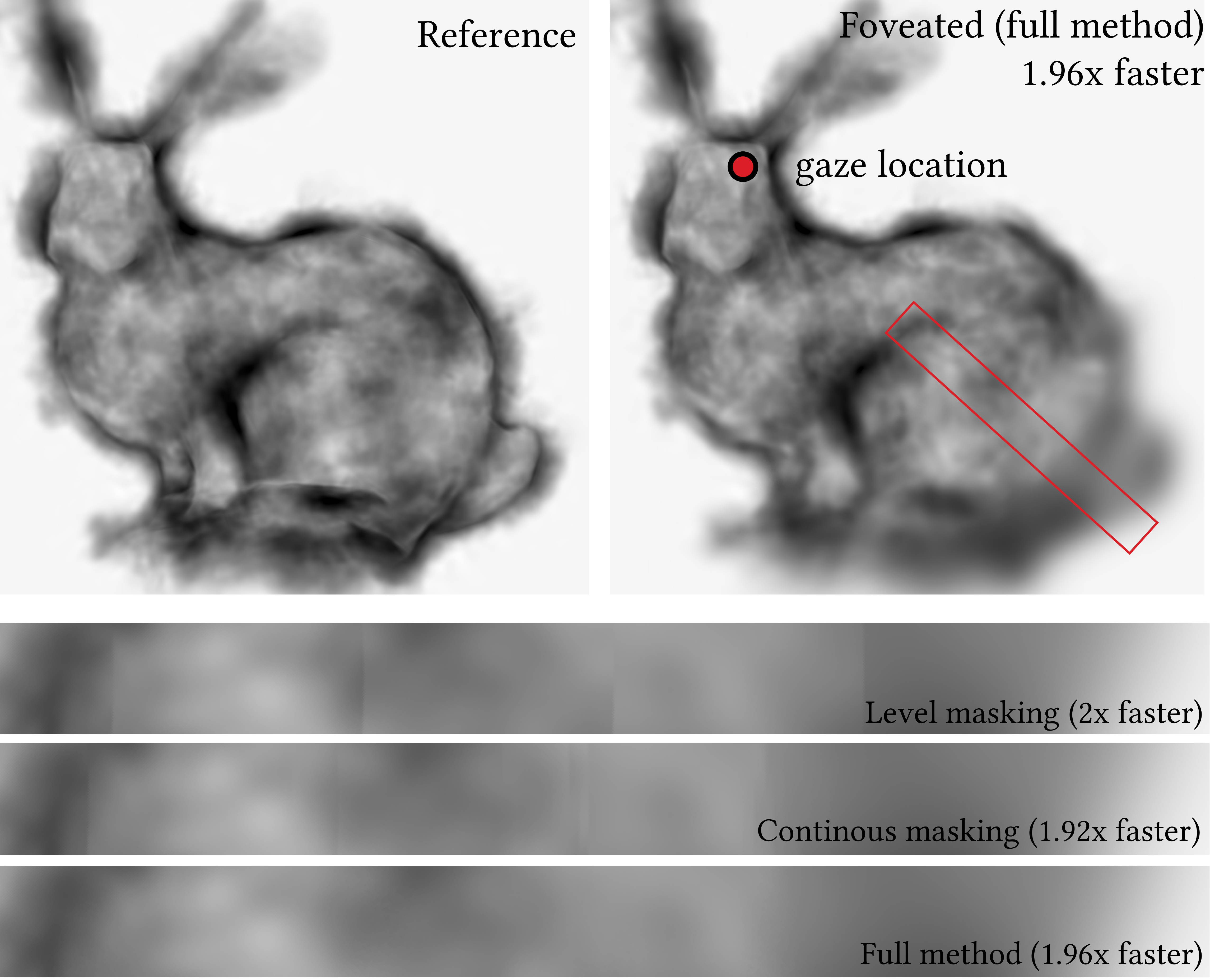}
    \caption{Results of proof-of-concept foveated renderering. The figure shows the full reference on the left and our foveated version on the right. The insets present different versions of our method with overall speedup with respect to the reference rendering. Note that under realistic foveated rendering conditions (i.e., following the CSF and proper eccentricity decay), level transitions in Level masking mode would be invisible.}
    \label{fig:foveation}
    \Description{Comparison of full reference rendering and foveated rendering versions with insets showing different levels of detail reduction and corresponding speedup factors for a volumetric cloud Bunny scene.}
\end{figure}

\subsection{Motion Blur}
Our representation provides an efficient way to filter primitives for motion-blur rendering. In particular, it can be used to discard primitives that contribute little to the motion-blur image at an early stage due to their frequency or orientation. One specific example is when a Gabor primitive moves half a cycle in the wave-plane direction, and the pair of original and shifted primitives cancel each other. This observation can be leveraged to reduce the cost of synthesizing motion blur, either by tuning sampling weights for specific directions and frequencies (Section~\ref{sec:rendering_gabor}) or by not considering primitives that contribute little to the final image. We implemented a basic motion blur using our method, which is further detailed in the Supplementary material.
\begin{figure*}
    \centering
    \includegraphics[width=\linewidth]{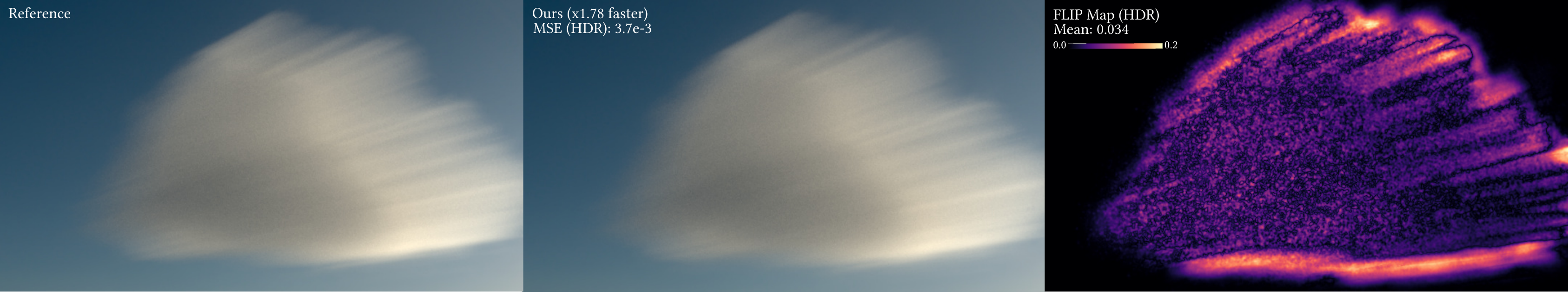}
    \caption{Results of our proof-of-concept motion blur implementation using the {\sc WDAS Cloud}. We compute the expected attenuation for a given linear motion (direction and magnitude), given an orientation and frequency decomposition of an asset (in this case, 3 frequencies and 3 orientations). At load time, we define ray visibility masks given this attenuation and a manually selected threshold, which results in culling two orientations at two frequency levels, amounting to a ~$23\%$ reduction in the number of particles. This renders a substantial performance uplift at minimal error, validating our attenuation model.}
    \label{fig:motion_blur}
    \Description{Comparison of motion blur renderings showing reference and accelerated versions with timing information, demonstrating culling of non-contributing orientations and frequencies for the WDAS Cloud dataset.}
\end{figure*}
\begin{figure}
    \centering
    \includegraphics[width=\linewidth]{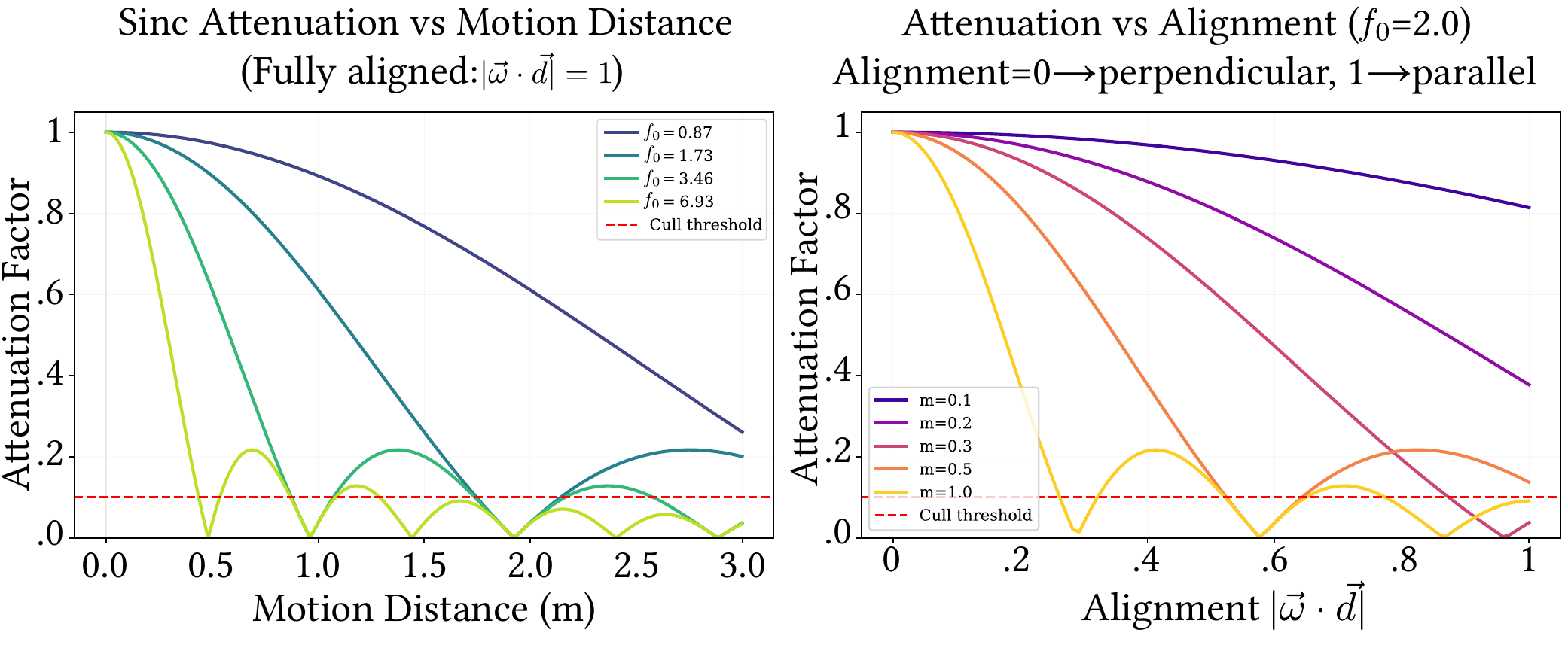}
    \caption{Behaviour of motion blur attenuation for different motion magnitudes, directions, and Gabor frequencies, and an arbitrary cutoff threshold.}
    \label{fig:sinc_attenuation}
    \Description{Plot showing sinc function attenuation curves for different motion parameters and Gabor frequencies, with a horizontal threshold line indicating culling cutoff.}
\end{figure}
Figure~\ref{fig:motion_blur} shows an example of applying motion blur to an asset.
We achieve a speedup of 1.78x over rendering the full asset, while obtaining almost the same result. This result is despite our Gabor kernels having low peak frequencies, which reduces their orientation selectivity.
This points out the practicality of our approach. Adapting directional importance sampling techniques could yield further efficiency gains and quality-speed trade-offs, as many combinations of motion directions and magnitudes result in significant attenuation. We consider this as an interesting extension. 
\subsection{Procedural Cloud Authoring Tool}
\begin{figure*}
    \centering
    \includegraphics[width=\linewidth]{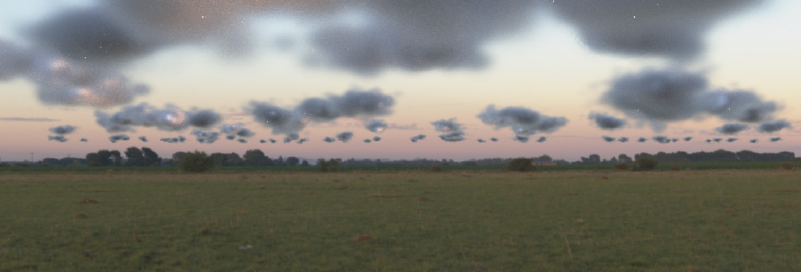}
    \caption{Rendering of 100 randomly placed, procedurally generated clouds. In total, the scene consists of 18902 Gaussian kernels and 308032 Gabor kernels. Clouds that are further back are entirely made up of Gaussian kernels. Cirrus clouds in the background are part of the environment map and not procedurally generated. }
    \label{fig:fields_gen}
    \Description{Wide outdoor scene showing 100 procedurally generated volumetric clouds of varying sizes and distances, rendered using Gabor and Gaussian kernels against a sky background with cirrus clouds.}
\end{figure*}

Inspired by the gradient noise density erosion common in industry~\cite{fajardo23procedural, Wrenninge2011}, we created our own authoring pipeline for cloud generation that leverages our volume representation. Through an interactive application, users can generate clouds that can be rendered directly with our approach without the need for any regression. To create a realistic cloud, we start by defining a cloud chunk similar to textured ellipsoids \cite{gardner1985visual}. 
In the Supplementary material, we detail how to generate these cloud chunks. Even though our cloud chunk can model the microstructure of a cloud, a typical cloud consists of several levels. A cloud can form a smaller cloud on its surface when inner pressure builds. To model this behavior, inspired by \citeauthor{bouthors2004modeling} \shortcite{bouthors2004modeling} and \citeauthor{dobashi2000simple} \shortcite{dobashi2000simple}, we define our cloud structure as a tree of cloud chunks. 
Figure \ref{fig:fields_gen} shows multiple procedurally generated clouds. Using the application the user can generate multiple clouds by controlling individual parameters and placing kernels in 3D space with their mouse to achieve any shape desired. A real-time tomography image of the generated cloud supports this process. Supplementary materials include a video of interaction with the application.

%% file: sections_paper/08_limitations_future_work.tex
\section{Limitations and Future Work}
\paragraph{Discussion on Gabor Transforms and Wavelets}
While the regression procedure delivers high-quality assets, it can be slow (0.1-2h depending on the asset and mixture complexity, on a single RTX 4090) and may require some small fine-tuning between assets of substantially different classes. We acknowledge that a forward method, akin to discrete wavelet decompositions (DWT) and Gabor Transforms, could be devised. However, Gabor kernels are not orthonormal, which generally rules them out as a valid wavelet basis; a complete frame without spectral overlap is impossible. While our work starts bridging the gap between wavelet and primitive-based rendering, we expect future work to finally enable seamless forward construction of kernel volumes without convolved regression schemes. An alternative approach for forward reconstruction could involve recent advances in Transformer-based feedforward point-based reconstruction approaches~\cite{wang2025vggt}.

\paragraph{Radiance fields}
While promising due to its similarity to cosine-modulated Gaussians, already showcased in 3DGRT~\cite{moenne20243dgrt} and Gabor filter banks for 2D images~\cite{wurster24gaborsplatting2d}, and despite some concurrent work showing limited success in similar radiance-field settings~\cite{zhou20253dgabsplat3dgaborsplatting,chan2026adagaradaptivegaborrepresentation}, there are fundamental limitations for using pure Gabor kernels (i.e., without a constant density base), mainly due to the view-dependence. Without dense enough training camera sampling, optimization could become tricky, as Gabor kernels might orient themselves into directions where they do not contribute to any image, appearing as artifacts after training. Regularization on the orientations, supervision on in-between frames \cite{Hermann_puzzlesim_iccv25} or better pruning strategies during training may improve its chances. Nevertheless, it remains a promising direction for future work, one which could enable a natural and continuous level of detail for primitive-based radiance field methods.

\paragraph{Limitations of Optix Visibility Masking}
While practical when using simple decompositions, using higher numbers of orientation bins and frequency levels would require changing the way we implement them, due to the limitation of 8 possible masks. Our alternative solution using the current Optix API, used in Section~\ref{sec:foveated_app}, is to define some through visibility masks, and others via direct masking in the rendering kernel (i.e., avoiding integration but keeping them in the visible set). In our experiments, the biggest bottleneck remains BVH traversal and ray intersection queries, which makes this latter strategy only marginally better in terms of performance.

\paragraph{Comparison to voxel grids} We include a comparison to a highly optimized voxel grid implementation in Mitsuba 3 in the Supplementary material. It includes hierarchical Super-Voxel (SV) structures~\cite{szirmay2011free}, tricubic texture sampling, a residual ratio tracking estimator~\cite{novak14} with local majorant estimates from the SV, and a weighted delta tracking distance sampler. Under these settings, our method, while achieving similar quality and unbiased rendering at a $\sim4300x$ compression rate, is substantially slower, despite also achieving smaller variance per sample, due to our analytic integration. This is a known trade-off between compression rate, quality and render time in primitive-based methods~\cite{Condor2024Gaussians}. Future work could explore different runtime optimization strategies, from hierarchical traversal~\cite{williams-2024}, complementary compression techniques (codebooks, half precision, fixed-frequency primitives), usage of anyhit shaders to collect overlapping primitives, decomposition tracking for distance sampling~\cite{kutz2017spectral}, and plenty of others.

\paragraph{Performance and Bias} While we manage to generally beat Gaussians across a wide range of metrics and datasets, in some scenarios (like extreme view-dependent frequency in the "volumetrically unrealistic" edges in {\sc{}Bunny}), the amount of residual kernels required to uplift a low-passed Gaussian version into a high-quality, high-frequency feature hampers performance substantially. Also, for multiple-scattering settings, most of our acceleration strategies rely on inducing certain amounts of bias to the render, which may not be adequate for all applications. 

\paragraph{Negative Density Fields} In practice, even when considering all overlapping primitives (both Gaussians and Gabors), some negative densities may occur on the volume’s boundaries. \emph{VPPT} can handle locally negative segments, due to the regular-tracking style integration. But when negative areas occur within a critical segment (i.e., the sampled one), the solver’s performance may decrease, due to breaking the assumption of a monotonically increasing function. In reality, the contribution of the solver is very small, and negative areas in the final mixture are very limited, which makes it a relatively small source of error.

\begin{figure}
    \centering
    \includegraphics[width=\linewidth]{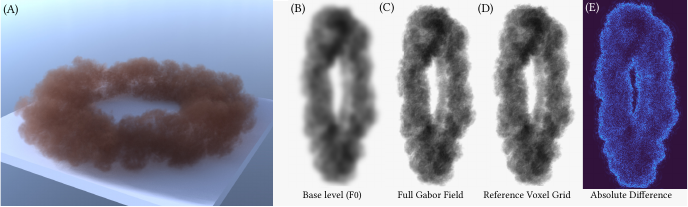}
    \caption{Multiple-scattering render of our {\sc{Dust}} asset (A); Tomographic visualization of its density field for the base, zero-frequency layer (B), the full Gabor field asset (C) and the original voxel grid reference (D); and absolute difference between the reference and the full Gabor field (E). When using aggressive low-pass filters for the lowest level, halo-like density artifacts can start to appear. This is due to it being a low-frequency error, compared to the high-frequency reference, which is attempted to be minimized through high-frequency residuals, requiring large quantities of Gabor kernels to remove; still, these are hardly noticeable in regular rendering settings (A).}
    \label{fig:halos}
\end{figure}

\paragraph{Halo Artifacts} In some scenarios, when the lowest frequency level modelled by base Gaussians (i.e., zero-frequency Gabor kernels) is very aggressive (as in, filtered with an extreme sigma), density halo-like artifacts can appear in the final asset. We visualize them in Figure~\ref{fig:halos}. Due to the extreme smoothness of the base layer, which drastically increases its spatial span, higher frequency residuals struggle to recover sharp edges, with a shallow density halo covering the asset. This is hardly visible in rendered assets, but becomes evident upon closer inspection. It essentially imposes a practical limit on the amount of filtering our method can efficiently model, as removing the halo would require larger numbers of residual primitives, straining the performance and compactness of the models.

%% file: sections_paper/09_conclusions.tex
\section{Conclusions}
We have presented Gabor Fields, a new volumetric kernel mixture. It naturally offers continuous frequency control by being tightly parameterized in the frequency domain, thus being perfect for LOD applications. We have provided analytical estimators for these new kernels and a specialized hierarchical regression technique. Furthermore, we have proposed a number of techniques that specifically leverage these primitives' properties to accelerate physically-based rendering, in particular exploiting angularity for the first time to significantly reduce scene traversal cost. We believe our work further connects procedural generation, wavelet volumetric approximations, and primitive-based volumes, providing a path towards ever more compact and efficient physically-based rendering. 

\paragraph{Acknowledgements}
This project has received funding from the Swiss National Science Foundation (SNSF, Grant 200502) and an academic gift from Meta. We acknowledge access to Alps at the Swiss National Supercomputing Centre, Switzerland under USI's share (project ID u6).

%% file: sections_paper/A_appendix.tex
\renewcommand{\thesubsubsection}{\thesection.\arabic{subsubsection}}

\section{Deriving Analytical Line Integrals of the 3D Anisotropic Gabor Kernel}
\label{sec:gabor_line_integral}
Here we present the step-by-step derivation of the closed-form definite line integrals for the 3D anisotropic Gabor kernel along an arbitrary ray.
Following the Gabor kernel defined in Eq.~\eqref{eq:gabor_kernel},

\paragraph{Kernel definition.}
Recall that
\begin{equation}
    \begin{aligned}
        g(x; \mu, \Sigma, \vec{\omega}) = \frac{1}{\sqrt{8\pi^3\left | \Sigma \right |}} e^{-\frac{1}{2}(x-\mu)^T\Sigma^{-1}(x-\mu)}\cos{\left (\vec{\omega}^T(x-\mu)\right )}
    \end{aligned}
\end{equation}
Defining the effective modulation vector
\begin{equation}
\vec{\omega}\;\equiv\; R S^{-1} (\omega,\omega,\omega),
\end{equation}
\paragraph{Line parameterization.}
Let the line be parameterized as
\begin{equation}
x(t) = p_0 + \vec{v}\,t, \qquad t\in[t_0,t_1],
\end{equation}%
and define the offset
\begin{equation}
d = p_0 - \mu.
\end{equation}
The line integral is
\begin{equation}
I
=
\int_{t_0}^{t_1} g(x(t))\,dt.
\end{equation}

\subsubsection{Reduction to a 1D Integral}
Substituting the line parameterization,
\begin{align}
I
&=
\frac{1}{\sqrt{8\pi^3 |\Sigma|}}
\int_{t_0}^{t_1}
\exp\!\left(
-\tfrac12 (d + t v)^T \Sigma^{-1} (d + t v)
\right)
\cos\!\left(
\vec{\omega}^T(d + t v)
\right)
dt.
\end{align}%
Introduce the scalar quantities, resulting from partial whitening of the space (only translation and rotation to local coordinates for better clarity)
\begin{align}
a &= v^T \Sigma^{-1} v, \\
\beta &= v^T \Sigma^{-1} d, \\
\gamma &= d^T \Sigma^{-1} d, \\
B &= \vec{\omega}^T v, \\
\delta &= \vec{\omega}^T d.
\end{align}%
Then
\begin{align}
(d + t v)^T \Sigma^{-1} (d + t v) &= a t^2 + 2 \beta t + \gamma, \\
\vec{\omega}^T(d + t v) &= B t + \delta.
\end{align}%
Thus,
\begin{equation}
I
=
\frac{1}{\sqrt{8\pi^3 |\Sigma|}}
\int_{t_0}^{t_1}
\exp\!\left(
-\tfrac12 (a t^2 + 2\beta t + \gamma)
\right)
\cos(B t + \delta)\,dt.
\end{equation}%

\subsubsection{Complex Exponential Form}

Using Euler's formula, we can transform the cosine as \(\cos\theta = \Re(e^{i\theta})\),
\begin{equation}
I
=
\frac{1}{\sqrt{8\pi^3 |\Sigma|}}
\Re\!\left\{
e^{i\delta}
e^{-\tfrac12 \gamma}
\int_{t_0}^{t_1}
\exp\!\left(
-\tfrac12 a t^2 - \beta t + i B t
\right)
dt
\right\}.
\end{equation}%
Completing the square,
\begin{equation}
-\tfrac12 a t^2 - \beta t + i B t
=
-\tfrac12 a
\left(
t - \frac{-\beta + i B}{a}
\right)^2
+
\frac{(-\beta + i B)^2}{2a}.
\end{equation}%

\subsubsection{Finite-Domain Closed Form}
Define
\begin{equation}
\alpha = \sqrt{\frac{a}{2}},
\qquad
\zeta = \frac{-\beta + i B}{\sqrt{2a}}.
\end{equation}
The finite-domain integral evaluates to
\begin{align}
I
&=
\frac{1}{\sqrt{8\pi^3 |\Sigma|}}
\Re\!\Bigg\{
e^{i\delta}
e^{-\tfrac12 \gamma}
\sqrt{\frac{\pi}{2a}}
\exp\!\left(
\frac{(-\beta + i B)^2}{2a}
\right)
\nonumber\\
&\hspace{3cm}
\times
\left[
\operatorname{erf}(\alpha t_1 - \zeta)
-
\operatorname{erf}(\alpha t_0 - \zeta)
\right]
\Bigg\}.
\end{align}%
This expression is exact and valid for any finite ray segment.

\noindent

\subsubsection{Infinite-Domain Limit}

Taking the limits \(t_0 \to -\infty\), \(t_1 \to +\infty\) and using
\(\operatorname{erf}(\pm\infty) = \pm 1\), we obtain
\begin{equation}
\boxed{
I_{\infty}
=
\frac{1}{\sqrt{8\pi^3 |\Sigma|}}
\sqrt{\frac{2\pi}{a}}
\exp\!\left(
-\tfrac12 \left(\gamma - \frac{\beta^2}{a}\right)
\right)
\exp\!\left(
-\frac{B^2}{2a}
\right)
\cos\!\left(
\delta - \frac{\beta B}{a}
\right).
}
\end{equation}%
In a fully whitened space (including scale), the expression simplifies by turning $a=1$, and multiplying by the corresponding length scaler to compensate for the space compression, rendering $\mathcal{K}$ as defined in Section 4.

\section{Frequency and Orientation Sampling Strategies} \label{sec:sampling_strats}
We include a number of stochastic-analytic estimators in Tables~\ref{tab:stochastic_sampling_strats} and~\ref{tab:steerable_sampling_strats}.
\begin{table*}[t]
  \centering
  \footnotesize
  \caption{Stochastic sampling strategies across frequency levels, with $\mathcal{P}$ being the number of levels in the pyramid, $\beta$ the power law shape parameter and $a, b$ the limits on the segment bucket the sample fell into (e.g. in a pyramid with $\mathcal{P} = 4$, the first level would have $a=0, b=\frac{1}{4}$). In control variate modes, Gaussians (i.e. zero-frequency Gabor kernels) have a weight of 1.}
  \label{tab:stochastic_sampling_strats}
   \begin{tabular}{@{}l l c c@{}}
    \toprule
    \textbf{Strategy} & \textbf{Description} & \textbf{Sampling Function $f(x)$} & \textbf{Sampling Weight (Reciprocal Probability)} \\
    \midrule
    Deterministic & Renders all frequency levels (no sampling). & -- & $1$ \\
    Uniform Sampling & Samples frequency levels uniformly. & $\mathcal{P}^{-1}$ & $\mathcal{P}$ \\
    Power-law Weighted & Samples proportionally to a tuned power-law function. & $x^{\left(1 - \beta \right)^{-1}}$ & $\left(b^{\left( 1-\beta  \right)} - a^{\left( 1 - \beta \right)}\right)^{-1}$ \\
    Uniform + Control Variate & Gaussians as a control variate; samples remaining levels uniformly. & $(\mathcal{P}-1)^{-1}$ & $\mathcal{P}-1$\\
    Power-law + Control Variate & Gaussians as a control variate; others sampled via power-law. & $x^{\left( 1 - \beta \right)^{-1}}$ & $\left(b^{\left( 1- \beta  \right)} - a^{\left( 1 - \beta \right)}\right)^{-1}$ \\
    Power-law + CV (Accum.) & Samples level $k$ via power-law; renders all levels $0$ to $k$. & $x^{\left( 1 - \beta \right)^{-1}}$ & $\left(1 - \left(\frac{j}{\mathcal{P}}\right)^{1-\beta}\right)^{-1}$ \\
    
    \bottomrule
  \end{tabular}
\end{table*}

\begin{table*}[t]
  \centering
  \footnotesize
  \caption{Orientation-selective sampling strategies for steerable pyramid rendering, with $K$ being the number of orientation bins, $\gamma$ the decay rate parameter, $a_i = |\vec{v} \cdot \mathbf{o}_i|$ the alignment between ray direction $\vec{v}$ and orientation bin $\mathbf{o}_i$, and $\delta$ the alignment threshold. The contribution weight is $w_i = \exp(-f_0^2 a_i^2/2)$, with $W = \sum_{i=1}^K w_i$. For deterministic threshold culling, we can simply tune "aggressiveness" in the form of the factor $\delta$; this allows us to sacrifice quality for performance in a direct manner, as lower thresholds result in more aggressive dropping of orientation bins, eventually degenerating into simply integrating the highest contributing one. Volumes with high $f_0$ will benefit from such a setting due to their high orientation selectivity, for example. Note: In Threshold + Control Variate mode, below-threshold bins have weight 1, and $N_{\text{above}}$ is the per-ray count of bins with $a_i > \delta$.}
  \label{tab:steerable_sampling_strats}
   \begin{tabular}{@{}>{\raggedright\arraybackslash}p{0.15\linewidth}>{\raggedright\arraybackslash}p{0.6\linewidth}>{\centering\arraybackslash}p{0.1\linewidth}>{\centering\arraybackslash}p{0.1\linewidth}@{}}
    \toprule
    \textbf{Strategy} & \textbf{Description} & \textbf{Sampling Probability} & \textbf{Sampling Weight} \\
    \midrule
    Deterministic & Renders all orientation bins (no culling). & -- & $1$ \\
    Threshold Culling & Renders bins with $a_i \leq \delta$ (perpendicular enough). & -- & $1$ \\
    Uniform Sampling & Uniformly samples one orientation bin. & $K^{-1}$ & $K$ \\
    Importance Sampling & Samples bins proportionally to contribution weight. & $\frac{w_i}{W}$ & $\frac{W}{w_i}$ \\
    Threshold + Uniform & Renders all bins with $a_i \leq \delta$; uniformly samples one with $a_i > \delta$. & $N_{\text{above}}^{-1}$ & $N_{\text{above}}$ \\
    \bottomrule
  \end{tabular}
\end{table*}

\section{Analytic Fourier Transform for Gabor Kernels}\label{sec:app_gaborfourier}
Recalling again that the Gabor kernel is defined as a Gaussian modulated by a cosine function:
\begin{equation}
    \begin{aligned}
        g(x; \mu, \Sigma, \vec{\omega}) = \frac{1}{\sqrt{8\pi^3\left | \Sigma \right |}} e^{-\frac{1}{2}(x-\mu)^T\Sigma^{-1}(x-\mu)}\cos{\left (\vec{\omega}^T(x-\mu)\right )}
    \end{aligned}
\end{equation}
In our experiments, we redefine $\vec{\omega}$ with a scalar $\omega$, but for the derivation, without loss of generalization, we use $\vec{\omega}$ as an arbitrary orientation of the modulation. We derive the analytic continuous Fourier Transform of the Gabor kernel using angular frequencies $k = 2 \pi \xi$. We define $y=x - \mu$ and use the shift theorem to shift to zero mean. 
\begin{equation}
    \begin{aligned}
        \mathcal{F}\{g(x)\}(k) = e^{-ik^T\mu} \mathcal{F}\{f(y)\}(k)
    \end{aligned}
\end{equation}
where
\begin{equation}
\begin{aligned}
    f(y) = \frac{1}{\sqrt{8\pi^3\left | \Sigma \right |}} e^{-\frac{1}{2}y^T\Sigma^{-1}y}\cos{\left (\vec{\omega}^Ty\right )}
\end{aligned}
\end{equation}
Using Euler's formula
\begin{equation}
    \begin{aligned}
        \cos(b^Ty) = \frac{1}{2} \left[e^{i\vec{\omega}^Ty} + e^{-i\vec{\omega}^Ty}\right]
    \end{aligned}
\end{equation}
and applying the frequency-shifting property
\begin{equation}
    \begin{aligned}
        \mathcal{F}\{h(y)e^{i \vec{\omega}^Ty} \}(k) = \mathcal{F}\{h(y)\}(k- \vec{\omega})
    \end{aligned}
\end{equation}
twice, we obtain by linearity
\begin{equation}
    \begin{aligned}
        \mathcal{F}\{f(y)\}(k) = \frac{1}{2}\left[\mathcal{F}\{h(y)\}(k-\vec{\omega}) + \mathcal{F}\{h(y)\}(k + \vec{\omega})\right]
    \end{aligned}
\end{equation}
where
\begin{equation}
    \begin{aligned}
        h(y) = \frac{1}{\sqrt{8\pi^3\left | \Sigma \right |}} e^{-\frac{1}{2}y^T\Sigma^{-1}y}
    \end{aligned}
\end{equation}
and
\begin{equation}
    \begin{aligned}
        \mathcal{F}\{h(y)\}(k) = e^{-\frac{1}{2}k^T\Sigma k}
    \end{aligned}
\end{equation}
Putting everything together, we conclude with the final form
\begin{equation}
    \begin{aligned}
       \mathcal{F}\{g(x)\}(k)  = \frac{1}{2} e^{-ik^T\mu} \left[e^{-\frac{1}{2}(k - \vec{\omega})^T\Sigma (k - \vec{\omega})} + e^{-\frac{1}{2}(k + \vec{\omega})^T\Sigma (k + \vec{\omega})} \right]
    \end{aligned}
\end{equation}
Lastly, we show how to convert between our continuous, angular frequency definition and Discrete Fourier Transforms (DFT) in cycles as computed by most FFT packages. Cycles and radians are related by $k=2\pi\xi$, thus $F_\text{cont}(k) = F_\text{cont}(2\pi\xi)$. Then we discretize our continuous function by approximating it with a  Riemann sum over a discrete grid $V \in \mathbb{R}^{N_x\times N_y \times N_z}$ and a physical domain $[a, b]^3$ with $a, b \in \mathbb{R}$ and a uniform sample spacing of $\Delta V = \Delta x \Delta y \Delta z = \frac{(b-a)^3}{N_x N_y N_z}$:
\begin{equation}
    \begin{aligned}
        \int_{[a,b]^3} f(\mathbf{r})e^{-2\pi i\xi^T \mathbf{r}}d\mathbf{r} \approx \Delta V\sum^{N_x - 1}_{n=0}\sum^{N_y - 1}_{m=0}\sum^{N_z - 1}_{l=0} f(\mathbf{r}_{n,m,l})e^{-2\pi i\xi^T \mathbf{r}_{n,m,l}}
    \end{aligned}
\end{equation}
The 3D DFT is unnormalized and defined as:
\begin{equation}
    \begin{aligned}
        F_\text{DFT}[p,q,r] = \sum^{N_x-1}_{n=0}\sum^{N_y-1}_{m=0}\sum^{N_z-1}_{l=0} f(\mathbf{r}_{n,m,l})e^{-2\pi i (\frac{pn}{N_x} + \frac{qm}{N_y}+\frac{rl}{N_z})}
    \end{aligned}
\end{equation}
thus
\begin{equation}
    \begin{aligned}
        \xi^T \mathbf{r}_{n,m,l} = \frac{pn}{N_x} + \frac{qm}{N_y}+\frac{rl}{N_z}
    \end{aligned}
\end{equation}
yields
\begin{equation}
    \begin{aligned}
        F_\text{cont}(2\pi \xi) \approx \Delta V F_\text{DFT}[p,q,r]
    \end{aligned}
\end{equation}
which we solve for $F_\text{DFT}$:
\begin{equation}
    \begin{aligned}
        F_\text{DFT}[p,q,r] = \frac{1}{\Delta V} F_\text{cont}(k_x = 2\pi\xi_p, k_y = 2\pi\xi_q, k_z = 2\pi\xi_r)
    \end{aligned}
\end{equation}
with $\xi_p = \frac{p}{N_x\Delta x}$ and $p = 0, ..., N_x-1$, and similarly for $\xi_q$ and $\xi_r$.

\section{Optimization Hyperparameters}\label{sec:app_hyperparams}

In Table~\ref{tab:hyperparameters}, we summarize the hyperparameters used to regress all assets presented in this work.

\newpage
 \input{tables/hyperparameters}

%% file: tables/hyperparameters.tex
\begin{table}[!ht]
\centering
\caption{Resulting optimization parameters after conducting hyperparameter searches on the {\sc bunny} volume.}
\begin{tabular}{>{\raggedright\arraybackslash}p{0.7\linewidth}>{\centering\arraybackslash}p{0.25\linewidth}}
\toprule
\textbf{Parameter} & \textbf{Learning Rate} \\
\midrule
\multicolumn{2}{l}{\textit{Parameter-specific Learning Rates}} \\
Centers $\mu$ & 0.012 \\
Scales $s$ & 0.0008 \\
Quaternions $q$ & 0.0012 \\
Weight $\alpha$ & 0.00008 \\
Omega $\omega$ & 0.002 \\
\midrule
\multicolumn{2}{l}{\textit{Regularization \& Control}} \\
$\lambda_{\alpha}$ & 0.00001 \\
$\lambda_{s}$ & 0.02 \\
$\lambda_\text{noise}$& 250 \\
$\gamma$ & 0.65 \\
\midrule
\multicolumn{2}{l}{\textit{Optimization Schedule}} \\
Warmup ratio & 0.15 \\
Min LR ratio & 0.1 \\
Iterations (Gaussian) & 300\\
Iterations (Gabor) & 300 \\
\midrule
\multicolumn{2}{l}{\textit{Densification \& Pruning}} \\
Resample Primitives & every 30 it \\
Resample until & 210 it \\
\bottomrule
\end{tabular}
\label{tab:hyperparameters}
\end{table}

%% file: sections_paper/B_supplementary.tex
\section{LOD Selection}
In the following, we show how to prefilter a Gabor mixture for LOD. Given a perspective camera with a field of view $\phi \; [rad]$ at resolution $R = \max(W, H)$ in pixels, we compute the visible width at distance $d \; [world]$ in world units as
\begin{equation}
    \begin{aligned}
        W_{\text{world}}(d) = 2d\tan{\left(\frac{\phi}{2}\right)}.
    \end{aligned}
\end{equation}
The angular Nyquist frequency in world units is then given by
\begin{equation}
    \begin{aligned}
        f_{\text{limit}}(d) = \frac{\pi R}{W_\text{world}(d)}.
    \end{aligned}
\end{equation}
We filter the mixture by comparing this cutoff to the peak frequency $f_0$ of each Gabor kernel (see Table 1 of the main paper). Note that using the peak frequency is an approximation, as no closed-form solution exists for the projection of a general Gabor kernel to screen space.
Counterintuitively, this threshold is independent of the screen-space extent of the mixture. Although larger objects cover more pixels, they also span proportionally more world-space distance, leaving the projected sampling density unchanged. Consequently, the cutoff depends only on camera parameters and depth, i.e., the local projection scale, rather than the overall size of the object in screen space.

\section{Comparison to Voxel Grids}
\begin{figure*}
    \centering
    \includegraphics[width=\linewidth]{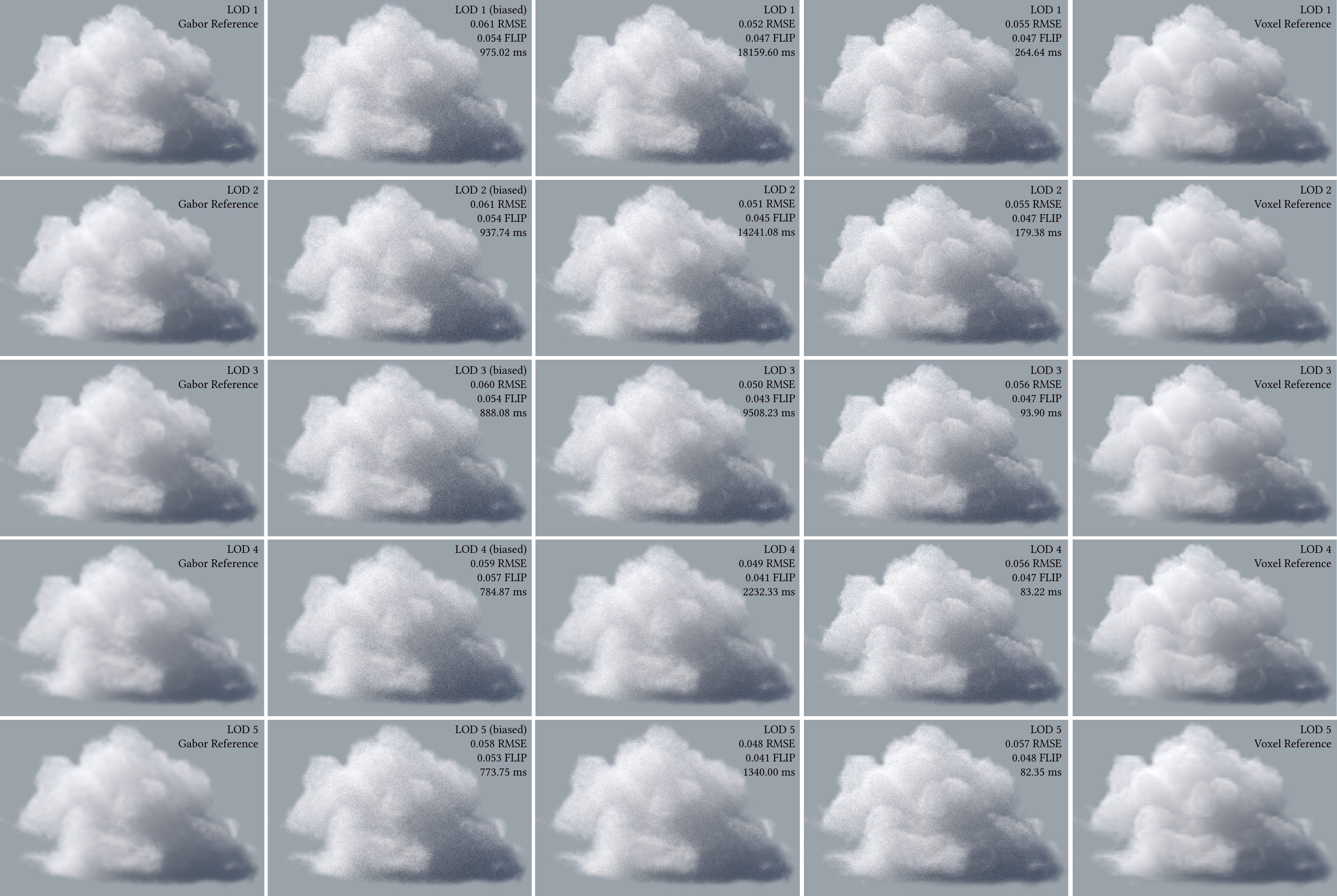}
    \caption{Multiple Scattering rendering comparison between Gabor fields and dense voxel grids + SuperVoxels~\cite{szirmay2011free}, using a residual ratio tracking~\cite{novak14} with local statistics. The first and last columns show converged renders and the middle columns show 8 spp renders. We report RMSE and FLIP between the converged reference LOD level and the lower spp counterpart. We used a 1.4x downsampled version of the {\sc{}WDAS Cloud} cloud as this represents the maximum voxel grid size that Mitsuba 3 supports. Subsequent levels are repeatedly downsampled by a factor of two (box filtering). Our mixture uses 46k primitives, yielding a $\sim{}4000x$ compression rate compared to the voxel grids. In the middle row, we showcase our unbiased integrator, while in the second row, we trade exactness for render time with a number of strategies: 1) after 3 recursions in the primary path, we only interact with the Gaussian level 2) use the Gaussian level for next event estimation shadow rays 3) uniform sampling of critical segments (no solver to invert the CDF) 4) adaptive clamping and 5) reduced number of maximum rounds for the Taylor series-based complex erf computation. All times were measured on an NVIDIA RTX 6000 Pro (Blackwell). }
    \label{fig:vox_comparisons}
    \Description{Multiple Scattering rendering comparison between Gabor fields and dense voxel grids + SuperVoxels~\cite{szirmay2011free}, using a residual ratio tracking estimator~\cite{novak14} with local statistics. The first and last columns show converged renders and the middle columns show 8 spp renders. We report RMSE and FLIP between the converged reference LOD level and the lower spp counterpart. We used a 1.4x downsampled version of the {\sc{}WDAS Cloud} cloud as this represents the maximum voxel grid size that Mitsuba 3 supports. Subsequent levels are repeatedly downsampled by a factor of two (box filtering). Our mixture uses 46k primitives, which results in a compression rate of $\sim{}4000x$ compared to the voxel grids. All times were measured on an NVIDIA RTX 6000 Pro (Blackwell).}
\end{figure*}

In Fig. \ref{fig:vox_comparisons}, we compare rendering performance between our Gabor fields and dense voxel grids. While both ran in Mitsuba 3, in order to make our comparison as fair as possible, we extended the native voxel grid support with the following:
\begin{itemize}
    \item SuperVoxels~\cite{szirmay2011free} for both faster distance sampling and providing aggregated statistics. At load time, the voxel grid is downsampled by a constant factor in each of its axes, and stored with aggregated medium density, local majorant and minorant statistics for the larger voxel.
    \item Residual Ratio Tracking~\cite{novak14} estimator for next-event shadow ray transmittance queries, using also aggregated statistics from the SuperVoxels structure.
    \item A weighted free flight distance sampler which also uses the local SuperVoxels to reduce memory traffic.
    \item Tricubic interpolation for texture taps to the volume grid, reducing "blocky" artifacts at lower levels of detail.
\end{itemize}
For each LOD, we showcase converged references next to low sample (8 spp) renderings and compute RMSE and FLIP between them. While voxel grids are significantly faster, our method yields lower variance per sample, thanks to analytic integration (middle column). We report unbiased results, but substantial acceleration can be obtained by employing stochastic integration strategies, adaptive clamping, and approximate segment sampling, which we showcase in the second column of Figure~\ref{fig:vox_comparisons} (described in more detail in Appendix B,  Figure 10 of the main paper, and the caption of Figure~\ref{fig:vox_comparisons}). Furthermore, our method achieves $\sim{}4000x$ compression rate vs the voxel grids, without using any particular compression scheme (e.g., quantization, compression, pruning), which could further increase the gap. We expect subsequent improvements to hardware support for primitive rendering, definition of primitive hierarchies for traversal, usage of anyhit shaders for overlap collection, or even employing decomposition tracking with analytical CDF inversion could soon bridge the performance gap.

\section{Foveated Rendering}
In the following, we discuss our implementation of foveated rendering with our method. We have implemented foveated rendering by assuming a linear relationship between eccentricity and the frequency threshold for discarding primitives. While correct foveation should consider the limitations of the human visual system \cite{strasburger-2011}, we focus here on demonstrating a proof-of-concept implementation. For each pixel, we first discard all levels that contain Gabor primitives with frequencies above the threshold. This step provides the most significant efficiency boost, as we can mask entire hierarchy levels, significantly reducing the cost of BVH traversal. Since the levels are discrete, this strategy does not provide truly continuous foveation. To improve continuity, we can check each remaining primitive's frequency content for the given ray direction and prevent its integration if its frequency falls below the threshold. The checks and additional code branching, however, make it slightly slower than pure level masking, although it depends on the asset and frequency distributions within each level. As the final step, we apply stochastic smoothing by randomly perturbing the per-pixel frequency threshold around transitions between levels.

\section{Motion Blur}
Here we share details about our motion blur implementation. In our simple motion blur implementation, we assume objects to move along the direction $\vec{d}$ with motion magnitude $m$. We model motion blur as a convolution with a 1D box filter oriented in direction $\vec{d}$ and with size $m$. In the frequency domain, this filter has a $\operatorname{sinc}$ function profile that attenuates frequencies aligned with the motion direction. For a Gabor kernel with modulation $\vec{\omega}$ this attenuation can be derived as:

\begin{equation}
    \text{attenuation} = \left|\operatorname{sinc}\bigl( m\:|\vec{\omega} \cdot \vec{d} | \bigr)\right|
\end{equation}
The alignment of $\vec{\omega}$ and $\vec{d}$ directly influences the attenuation. As the vectors get more parallel to each other, $|\vec{\omega} \cdot \vec{d}|$ approaches $f_0=\|\vec{\omega}\|$, the motion causes destructive interference, especially for small Gabor primitives. Conversely, when the vectors become perpendicular, the primitives are not attenuated and fully contribute to the resulting image. Note that this analysis holds for any motion in 3D regardless of the camera orientation. We present the analysis of the attenuation function in Figure 14 of the main paper.

To implement this strategy, we first group Gabor primitives into bins $\boldsymbol{B}_i$ containing different orientations and frequencies. Given motion parameters $(\vec{d}, m)$, we cull bin $i$, if the attenuation for the mean frequency and orientation in that bin falls below a fixed threshold.

\section{Procedural Cloud Generation}
In the following sections, we detail our strategy for procedural generation of clouds. A cloud chunk is defined using a Gaussian kernel and several Gabor kernels on its surface. This way, we can represent the core of a cloud and the splittings and wispy look caused by effects such as wind. For the placement of Gabor kernels, we first define a grid on the surface. Similar to Gabor noise, a Gabor kernel is generated within a random position inside each grid cell \cite{galerne12, lagae2009procedural}. The frequency and the orientation of these Gabor kernels are randomly picked from a uniform distribution. Scales of the kernels, on the other hand, are randomized using a Maxwell distribution \cite{wu2018statistical}, which provides a physical intuition for the stability of the cloud.

When the cloud collides with a higher temperature upstream, it encounters water bubbles with increased speed. This results in higher inner pressure for the cloud. When the inner pressure breaks through the outer edge of a cloud, a smaller cloud that is semi-connected to the original cloud develops. Keeping this in mind, we used our cloud chunks to generate more complex cloud formations. A full cloud is generated through randomly generated cloud chunks in layers similar to a fractal. These multi-layer clouds have smaller cloud chunks on their surface. These smaller chunks are placed uniformly at random on the surface, with the radius drawn from a Maxwell distribution.

Another complex cloud formation occurs when two different clouds merge. This is easily doable in our application by generating different multi-layer clouds and merging them by moving them together.